\documentclass[
aps,showpacs
 ,twocolumn
]{revtex4}
\usepackage{tikz}
\usepackage{adjustbox}
\usetikzlibrary{arrows}
\usepackage{wrapfig}
\usepackage{graphicx}
\usepackage{amsmath}
\usepackage{amssymb}
\usepackage{hyperref}
\usepackage{ mathrsfs }
\usepackage{indentfirst}
\usepackage{braket}
\usepackage{color}
\usepackage{empheq}
\usepackage{lipsum,adjustbox}
\usetikzlibrary{calc,positioning}

\begin{document}

\title{Estimating space-time parameters with a quantum probe in a lossy environment}

\author{ S. P. Kish and T. C. Ralph}\affiliation{
School of Mathematics and Physics, University
of Queensland, Brisbane, Queensland 4072, Australia}

\date{\today}

\begin{abstract}
{}
We study the problem of estimating the Schwarzschild radius of a massive body using Gaussian quantum probe states. Previous calculations assumed that the probe state remained pure after propagating a large distance. In a realistic scenario, there would be inevitable losses. Here we introduce a practical approach to calculate the Quantum Fisher Informations (QFIs) for a quantum probe that has passed through a lossy channel. Whilst for many situations loss means coherent states are optimal, we identify certain situations for which squeezed states have an advantage. We also study the effect of the frequency profile of the wavepacket propagating from Alice to Bob. There exists an optimal operating point for a chosen mode profile. In particular, employing a smooth rectangular frequency profile significantly improves the error bound on the Schwarzschild radius compared to a Gaussian frequency profile. 
\end{abstract}


\maketitle


\vspace{10 mm}
\section{Introduction}

The precision with which physical parameters can be estimated is limited by the level of fluctuations or noise in the measurement device. In general, quantum mechanics introduces irreducible levels of noise onto measurement results, and hence quantum mechanics places limits on the ultimate precision of parameter estimation. The study of these limits and the development of protocols for reaching them is called quantum metrology \cite{GIO11}. In optics, the use of semi-classical probe states such as coherent states, where the quantum noise can be interpreted as photon shot-noise, leads to a standard quantum limit. To surpass this limit requires the use of non-classical states displaying squeezing or entanglement \cite{milburn}.

Most quantum metrology assumes non-relativistic quantum mechanics. However, many applications of quantum metrology relate to relativistic phenomena such as gravitational waves \cite{GRO13} and the estimation of gravitational fields and accelerations \cite{SOR11}. More rigorous approaches to such problems, that will become important as precision grows, use relativistic quantum field theory to describe the quantum interactions in spacetime \cite{birrell}. A number of authors have begun exploring such approaches \cite{DOW11, AHM14, AHM14a}. 
 
 \tikzstyle{int}=[draw, fill=blue!20, minimum size=2em]
\tikzstyle{init} = [pin edge={to-,thin,black}]
\begin{figure*}
\begin{adjustbox}{width=\textwidth}
\begin{tikzpicture}[node distance=7.0cm,auto,>=latex']
     \draw[black,dashed] (-5,-1.5) -- (-1.75,-1.5) -- (-1.75,1.5) -- (-5,1.5) node[pos=0.5]{Pure probe state}-- (-5,-1.5);
     
     \draw[red,dashed] (-1.5,-1.5) -- (5.5,-1.5) -- (5.5,1.5) -- (-1.5,1.5) node[pos=0.5]{Lossy probe state}-- (-1.5,-1.5);
     
     \draw[blue,dashed] (5.75,-1.5) -- (8,-1.5) -- (8,1.5) -- (5.75,1.5) node[pos=0.5]{Parameter}-- (5.75,-1.5);
     
     \draw[green,dashed] (8.25,-1.5) -- (12.5,-1.5) -- (12.5,1.5) -- (8.25,1.5) node[pos=0.5]{Unitarily evolved state}-- (8.25,-1.5);
    \node [int, pin={[init]below:$X_{c}=c+c^{\dagger}$}] (a) {$\hat U_{BS_t}$};
    \node (b) [left of=a,node distance=5cm, coordinate] {a};
    \node [int, pin={[init]below:$X_{d}=d+d^{\dagger}$}] (c) [right of=a] {$\hat U_{BS_\Theta}$};
    \node [coordinate] (end) [right of=c, node distance=6cm]{};
    \path[->] (b) edge node {$X_{in}=a+a^{\dagger}$} (a);
    \path[->] (a) edge node {$X_{b}=t X_{in}-\sqrt{1-t^2}X_{c}$} (c);
    \draw[->] (c) edge node {$  X_{out}=\Theta X_{b}-\sqrt{1-\Theta^2} X_{d}$} (end) ;
\end{tikzpicture}
\end{adjustbox}
\caption{Representation of a quantum channel. A pure Gaussian probe state passes through a lossy channel of transmission $t$. It is equivalent to the probe state evolving under the unitary beamsplitter operator $U_{BS_t}$. The subsequent `lossy probe state' will be used to measure the beamsplitter parameter $\Theta$ of the unitary operator $U_{BS_\Theta}$.}
\label{qchan}
\end{figure*}
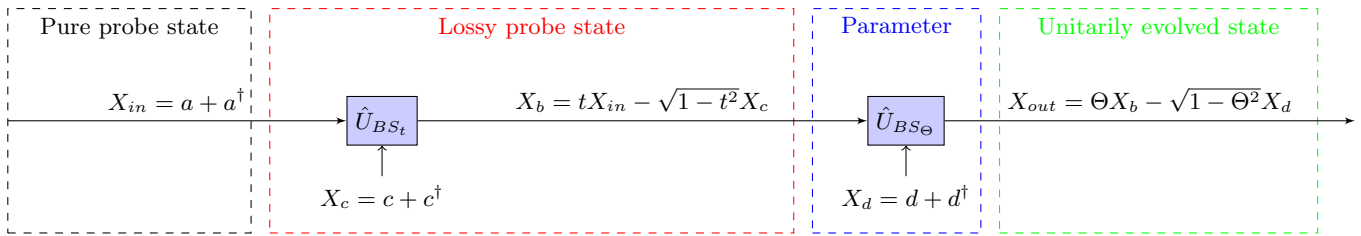
 
Recently Bruschi et al \cite{BRU14} showed that techniques for optimally estimating the transmission parameter of a quantum channel \cite{MON07} can be adapted to the relativistic problem of estimating the Schwarzschild radius of a massive body. Their protocol involves coherently comparing a quantum optical probe state prepared at one height in the metric with a second, locally identical probe state, prepared at a different height. They investigated the use of optical probes prepared in coherent states and squeezed states and found that using squeezed vacuum states was optimal. However, the calculations of Ref.\cite{BRU14} assumed that losses could be neglected, even though the probe states potentially needed to be propagated over large distances in the protocol. In addition, only Gaussian temporal wave-packets were considered and a limited region of the parameter space was explored.

In this paper we analyse a more realistic version of the Bruschi et al protocol that includes the inevitable losses that would occur in such a scheme in practice and optimises the parameters with respect to height, bandwidth, mode-shape and operating point. We find that squeezing is not always optimal but can enhance precision under certain conditions. In our analysis we introduce a different approach to obtaining the Fisher information for this protocol which turns out to be much easier to generalize to more realistic scenarios.

The paper is arranged in the following way. In the next section we review the basic principles of estimating the transmission parameters of a quantum channel using a quantum probe, describing our approach to obtaining the relevant quantum Fisher informations and calculating results for mixed probe states. In section III we review how this approach can be adapted to estimating parameters associated with space-time curvature, in particular the Schwarzschild radius, $r_s$, of a massive body. In section IV we apply our formalism to this problem and derive expressions for the relative errors in estimating $r_s$ in a number of idealised scenarios. We make more realistic assumptions in section V, for example incorporating loss as a function of transmission distance and considering bright coherent beams with added squeezing. We conclude and discuss in section V.

\section{Transmission of a quantum channel}

The goal of any quantum estimation is to determine a probe state and probability operator-value measurement (POVM) represented by the estimator $\hat {\Theta}$ and to determine the value of $\Theta$ from the set of $N$ measured outcomes \cite{MON07}. We consider an unbiased estimator such that for $N \to \infty$, the expectation value $E[\hat \Theta]$ returns $\Theta$ and all errors disappear. The bound for the variance of an unbiased estimator is set by the Cramer-Rao inequality \cite{cramer}. 

\begin{equation}
\braket{\Delta \hat \Theta ^2} \ge \frac{1}{F(\Theta)} 
\label{cramer}
\end{equation}

Where $F(\Theta)$ is the Fisher information of a measurement. The Fisher information coincides with the second moment of the classical logarithmic derivative of the likelihood function. For $N$ number of measurements of identically prepared quantum states, the total Fisher information is the sum of all individual Fisher informations. Therefore, this implies that the variance of a parameter scales as $\frac{1}{NF(\Theta)}$. The Fisher information $F(\Theta)$ is further bounded by the {\it Quantum Fisher information} $H(\Theta)$ which signifies the most precise measurement allowed by quantum mechanics. Similarly, the QFI is additive and for $N$ measurements we have the bound:

\begin{equation}
\braket{\Delta \hat \Theta ^2} \ge \frac{1}{N F(\Theta)} \ge \frac{1}{N H(\Theta)} 
\label{cramer}
\end{equation}

The general problem we wish to address in this section is to determine the Cramer-Rao bound for the beamsplitter parameter $\Theta$ if the probe states initially evolve under a lossy quantum channel given by a known transmission $t$. We represent the situation diagrammatically in Fig. 1. It is always possible to decompose the lossy channels into orthogonal modes. We begin by introducing an auxiliary mode and treating the loss and parameter estimator as two beamsplitters with transmissions $t$ and $\Theta$.   

We consider the beamsplitter transformations,
\begin{eqnarray}
\hat b^{\dagger} = t \hat {a}_{in}^{\dagger}-\sqrt{1-t^2} \hat c^{\dagger} \\
\hat a_{out}^{\dagger} = \Theta \hat{b}^{\dagger}-\sqrt{1-\Theta^2} \hat d^{\dagger}
\label{e1}
\end{eqnarray}

Where $\hat a_{in}$ is Alice's input mode and $\hat b$ is the mode Bob receives. The auxiliary modes are $\hat c_{out}$ and $\hat d_{out}$. Our goal is to determine an appropriate {\it probe state} for $\hat a_{in}$ that maximises the QFI under the evolution of the {\it lossy quantum channel}. The QFI can be written as the symmetric logarithmic derivative (SLD) defined as a Hermitian operator that has the form \cite{MON07}:

\begin{equation}
\frac{ d \hat \rho_{\Theta}}{d \Theta}=\frac{1}{2} [\hat \rho_{\Theta} \hat \Lambda (\Theta) + \hat \Lambda (\Theta) \hat \rho_{\Theta}]
\end{equation}

Where $\hat \Lambda (\Theta)$ is an optimal system observable with an expectation value $Tr [\hat \Lambda (\Theta) \hat \rho_{\Theta}]=0$, and $\rho_{\Theta}$ is the density operator describing the output state of the probe. In order to determine the QFI, we can evaluate the operator $\Lambda( \Theta)$ and thus $H(\Theta)=Tr [\hat \Lambda^2 (\Theta) \hat \rho_{\Theta}]$ as done in Ref \cite{MON07}. 

However, we consider a more practical and succinct approach to determine the QFI that is useful if the additional known loss parameter $t$ is introduced and the probe state is Gaussian. We assume that $t$ is completely characterised beforehand. The representation of this lossy quantum channel in Fig. \ref{qchan} consists of a two beamsplitter setup with $t$ and $\Theta$ corresponding to the transmissions. A {\it pure Gaussian probe state} passes through a lossy channel of transmission $t$ represented by the probe state evolving under the unitary beamsplitter operator $U_{BS_t}$. The subsequent `lossy probe state' will be used to measure the beamsplitter parameter $\Theta$ of the unitary operator $U_{BS_\Theta}$. To determine the QFI, we begin from the geometry of two density matrices.
The Bures distance is the minimal distance between purifications of two density matrices $\rho$ and $\sigma$:
\begin{equation}
d_B(\rho, \sigma)=[2 (1-\sqrt{\mathcal{F}(\rho,\sigma)}]^{\frac{1}{2}}
\end{equation} 

Where $\mathcal{F}(\rho,\sigma)$ is the quantum fidelity:
\begin{equation}
\mathcal{F}=(Tr(\sqrt{\sqrt{\rho} \sigma \sqrt{\rho}}))^2
\end{equation}

The quantum fidelity or Uhlmann's transition probability is a well known quantification \cite{jeong} for the similarity of quantum states. The Bures distance can be related to the quantum Fisher information via:

\begin{equation}
H(\Theta)=\lim_{d\Theta \to 0} \frac{8(1-\sqrt{\mathcal{F}(\rho_{\Theta},\rho_{\Theta+d\Theta})})}{d\Theta^2}
\label{qfib}
\end{equation}

For pure states, the fidelity reduces to the overlap of the initial and final states $\mathcal{F}=|\braket{\psi|\psi'}|^2$. For a general Gaussian state of any mixedness, the fidelity can be expressed in terms of the quadrature variances $V^{+}=\braket{\Delta X(\phi)^2}$ and $V^{-}=\braket{\Delta P(\phi)^2}$, where $X=a + a^{\dagger}$ and $P=-i(a -a^{\dagger})$. The variances $V^{\pm}$ are directly measurable values. We wish to determine the final quadrature variances of the evolved state $V_1^{\pm}$ for a parameter value $\Theta$, and the variance $V_2^{\pm}$ for an infinitesimal change $\Theta+d\Theta$. The fidelity can be expressed in the form \cite{jeong, twa, wang}:

\begin{equation}
\mathcal{F}=\mathcal{F}(\phi_s) \mathcal{D}(x)
\end{equation}

$\phi_s$ is the angle between the two states. We assume that for the rest of this paper that this is unchanged $\phi_s=0$. $\mathcal{F}$ at $\phi_s=0$ can be expressed in terms of the quadrature variances $V_1$ and $V_2$:

\begin{equation}
\begin{split}
\mathcal{F}(\phi_s=0)&= 2 \{ \sqrt{(V_1^{+} V_2^{-}+1)(V_1^{-} V_2^{+}+1)} \\
&-\sqrt{(V_1^{+}V_1^{-}-1)(V_2^{+}V_2^{-}-1)} \} ^{-1}
\end{split}
\end{equation}

In addition, if the two states are separated by $x_r+ix_i$ in phase space, a factor $\mathcal{D}(x)$ is introduced:

\begin{equation}
\mathcal{D}(x)=\exp[- \frac{2 x_r^2}{V_1^++V_2^+}- \frac{2 x_i^2}{V_1^-+V_2^-}]
\end{equation}

\subsection{Coherent probe state with thermal noise}

We can use these expressions to determine the QFI for a {\it coherent probe state} with a {\it mixed thermal state} in the auxiliary mode $\hat c$. The variances of the quadratures are obtained from the Heisenberg picture. The variances add as follows $\braket{(\Delta X_{out})^2}=t^2 \braket{(\Delta X_{in})^2}+ (1-t^2) \braket{(\Delta X_{c})^2}$ and $\braket{(\Delta X_{b})^2} =\Theta^2 \braket{(\Delta X_{out})^2}+ (1-\Theta^2) \braket{(\Delta X_{d})^2}$. A thermal state has variance $\braket{(\Delta X_{c})^2}=2\tilde{n}_{Th}+1$ and $\braket{(\Delta X_{d})^2}=1$ is the vacuum state. Hence

\begin{equation}
V_1^{+}=2 \tilde{n}_{Th} \Theta^2 (1-t^2)+1
\end{equation}
Where $\tilde{n}_{Th}$ is the Planck thermal distribution function.
Thus for a slight variation in the parameter $\Theta+d\Theta$, the variance is:

\begin{equation}
V_2^{+}=V_1^{+}+4 \tilde{n}_{Th} \Theta d \Theta (1-t^2) +2 \tilde{n}_{Th} d\Theta^2 (1-t^2)
\end{equation}

It can be shown that $V_1^{-}=V_1^{+}=V_1$ and $V_2^{-}=V_2^{+}=V_2$. Thus the equation for fidelity (5) is reduced to:

\begin{equation}
\mathcal{F}(\phi_s=0)=2 \{ (V_1 V_2+1) -\sqrt{(V_1^2-1)(V_2^{2}-1)} \} ^{-1}
\end{equation}

Thus the total fidelity to second order in $d \Theta$ is:
\begin{equation}
\begin{split}
\mathcal{F}&=1- \frac{d\Theta^2 |t\alpha|^2}{V_1}- 4 \frac{d \Theta^2 \tilde{n}_{Th}^2 \Theta^2 (1-t^2)^2 }{(V_1^2-1)}\\
&=1-  \frac{d\Theta^2 |t\alpha|^2}{2\tilde{n}_{Th}\Theta^2 (1-t^2)+1}- \frac{d\Theta^2 \tilde{n}_{Th} (1-t^2) }{\tilde{n}_{Th}\Theta^2(1-t^2)+1}
\end{split}
\end{equation}

We make use of the definition in equation \ref{qfib} and the binomial expansion for $x<<1$, $\sqrt{1-x}\approx 1-x/2$. Finally, the QFI is given by:

\begin{equation}
\begin{split}
H(\Theta)&=\frac{4 |t\alpha|^2}{V_1}+ \frac{16 \tilde{n}_{Th}^2 \Theta^2 (1-t^2)^2 }{(V_1^2-1)}\\
&= \frac{4 |t\alpha|^2}{2\tilde{n}_{Th}\Theta^2 (1-t^2)+1}+ \frac{4 \tilde{n}_{Th} (1-t^2) }{\tilde{n}_{Th}\Theta^2(1-t^2)+1}
\end{split}
\label{ntherm}
\end{equation}

We note that for room temperature $T=300$ K and signal frequency $\omega=700$ THz ($\lambda=430$ nm), the thermal number occupation $\tilde{n}_{Th}$ is negligible. Thus, the QFI can be approximated to that of an attenuated coherent state: 

\begin{equation}
H(\Theta)=4 |t\alpha|^2
\label{coh}
\end{equation}

\subsection{Squeezed Coherent probe state}

We now consider a {\it squeezed coherent probe state}. We assume the auxiliary vacuum states in either beamsplitters have variance $\Delta X_{c}=\Delta X_{d}=1$. A squeezed coherent state can have a quadrature variance that is better than the shot noise $\braket{(\Delta X_{in})^2}=e^{-2r}$ and $\braket{(\Delta P_{in})^2}=e^{2r}$. The level of squeezing is determined by the parameter $r$ where we have assumed the maximum squeezing without loss of generality is in the $X$ quadrature. We note that the squeezing parameter $r$, the magnitude $|\alpha|$ and the angle $\theta$ of the coherent state are the only relevant parameters Thus, the variances of the evolved state are $V_1^{+}= \Theta^2 t^2 (e^{-2 r}-1)+1$ and $V_1^{-}= \Theta^2 t^2 (e^{2 r}-1)+1$. Since we are estimating how well a change in $\Theta$ can be detected, the second state is the same state with an infinitesimal shift in the parameter $\Theta+d \Theta$. In phase space, the separation is given by $x_r+ix_i=d\Theta t( \cos(\theta)+i \sin(\theta)) |\alpha|$. We approximate the fidelity expression to second order in $d \Theta$ and disregarded any higher orders. 

\begin{figure}[tp]
\includegraphics[width=1.12 \linewidth]{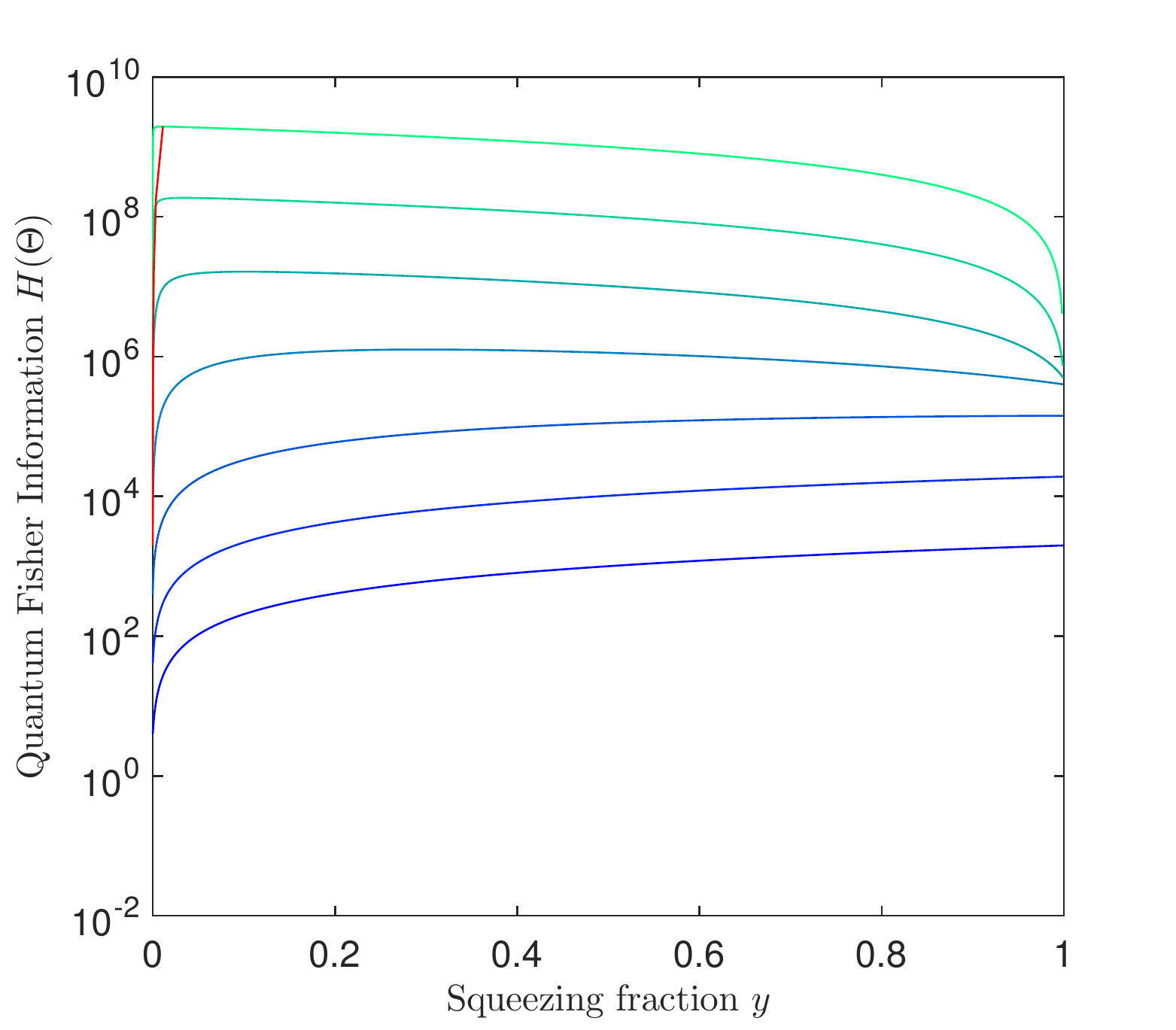}
\caption{The QFI for perfect transmission $t=1$ for various probe energies $\tilde{n}$. For $\tilde{n} \ge 10^2$, it is no longer advantageous to use a large fraction of squeezed photons. The parameter to be measured is chosen to be $\Theta= 1-1.0\times 10^{-3}$.}
\label{fig2}
\end{figure}   
   
 \begin{figure}
\includegraphics[width=1.12 \linewidth]{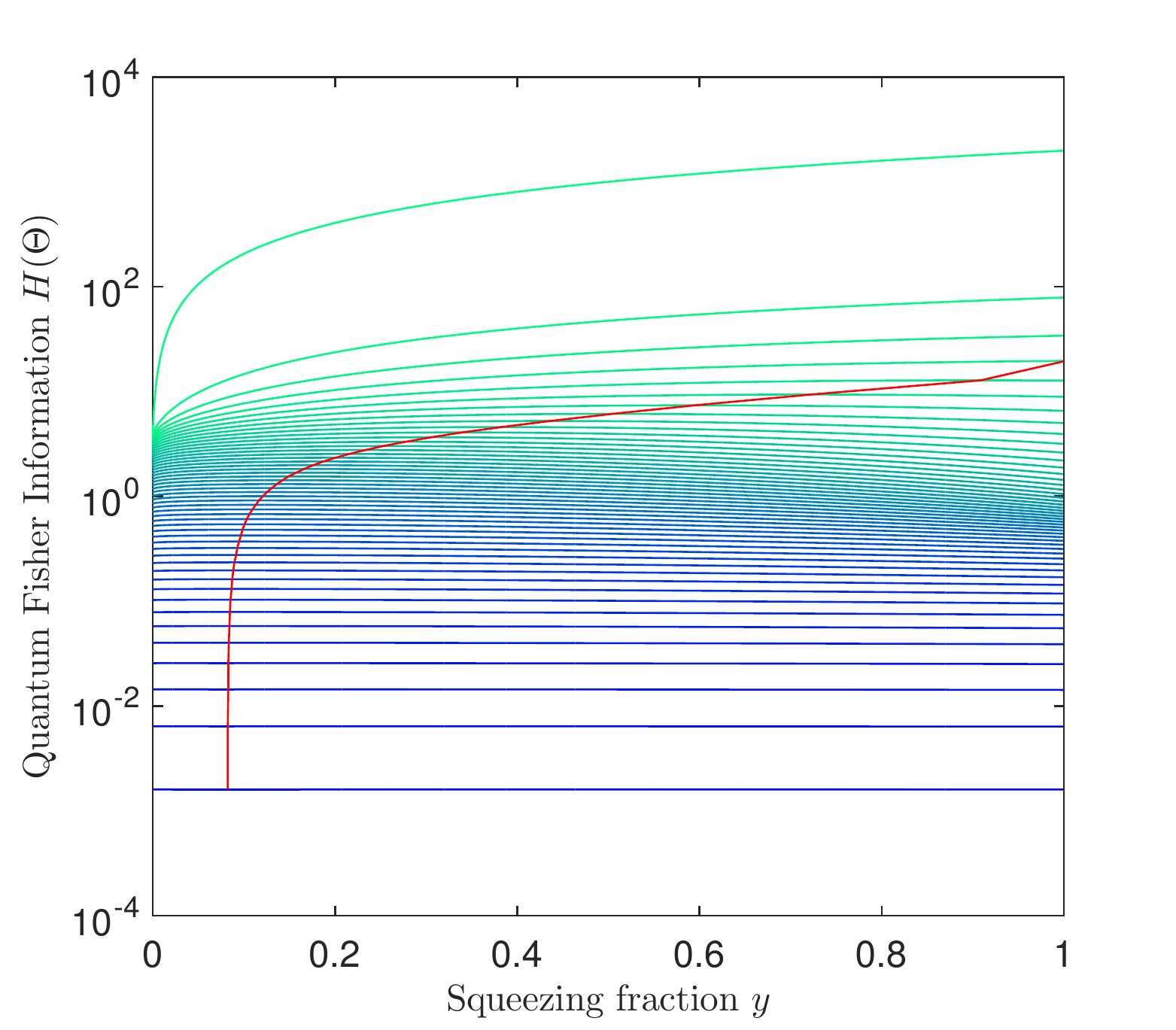}
\caption{The Quantum Fisher Information for a squeezed coherent state with probe state energy $\tilde{n}=1$. The channel transmission $t$ varies from $t=0$ (blue) to $t=1.00$ (green) in $0.02$ intervals. The red curve represents the maximum QFI for each $t$. The parameter is $\Theta=1-x \approx 1-1.0 \times 10^{-3}$}
\label{onen}
\end{figure}

\begin{figure}
  \includegraphics[width=1.12 \linewidth]{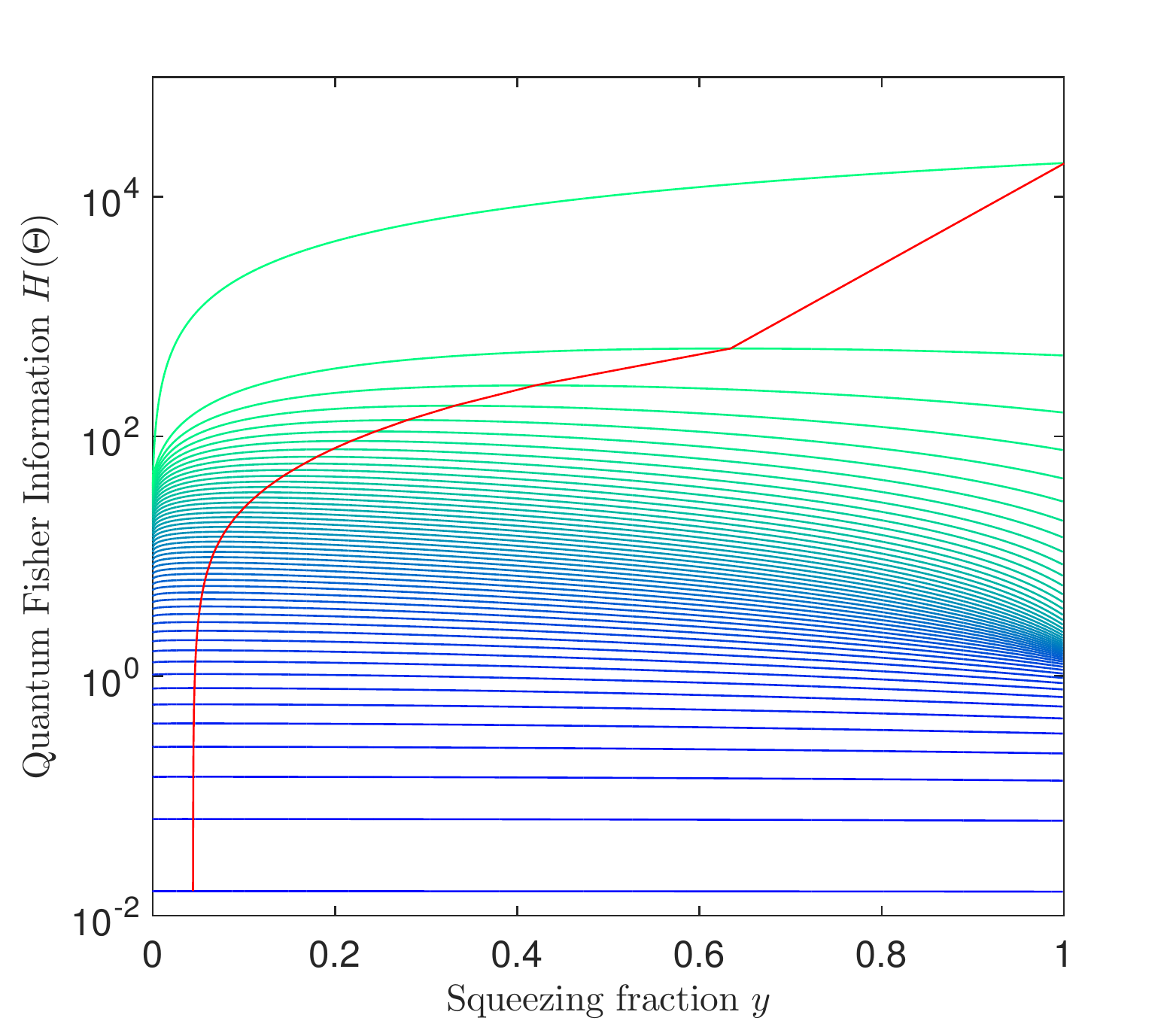}
  \caption{The QFI for a squeezed coherent state with probe state energy $\tilde{n}=10$. Same $\Theta$ as in Fig. \ref{onen}}
   \label{okay}
\end{figure}

\begin{figure}
  \includegraphics[width=1.12 \linewidth]{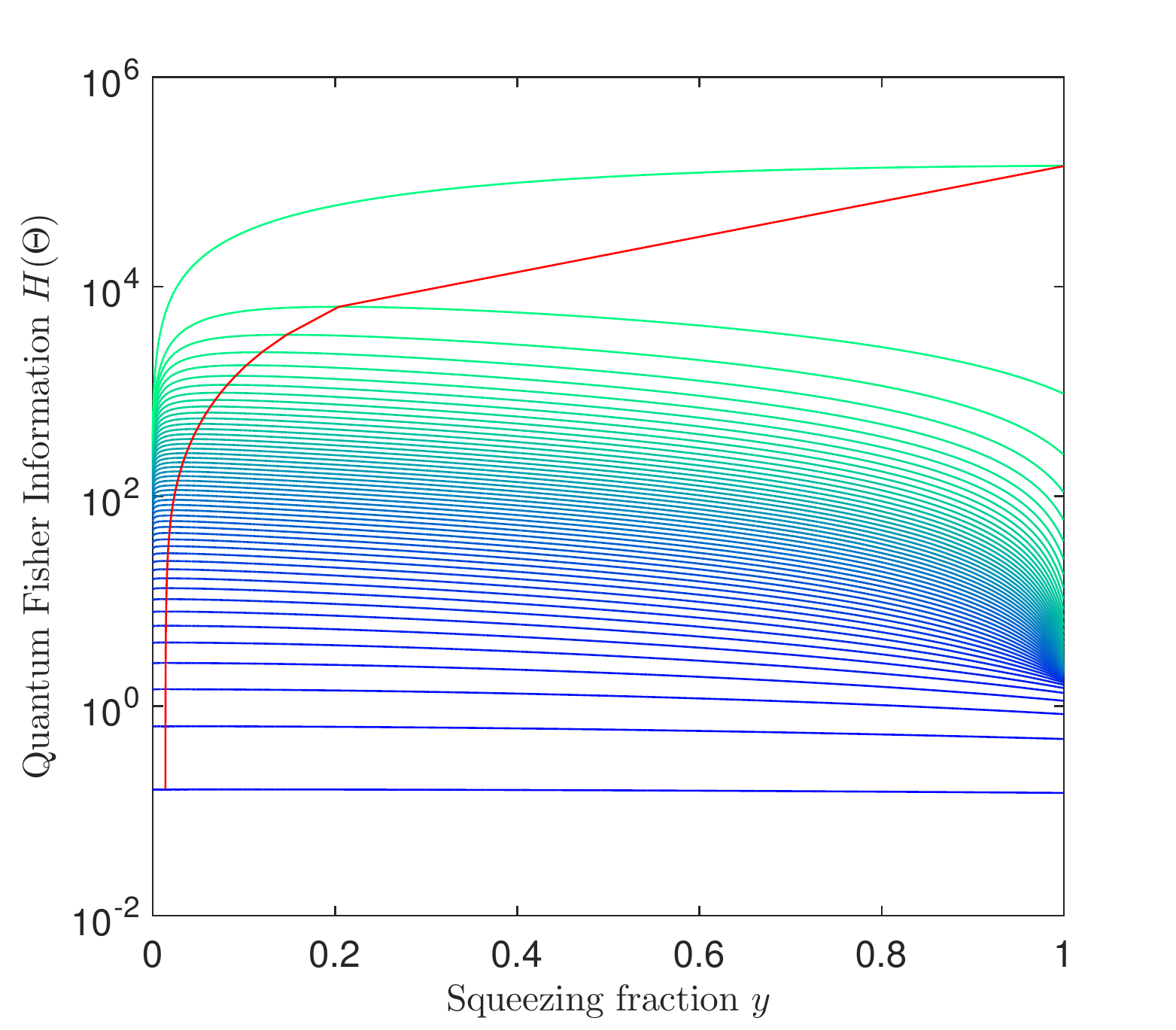}
  \caption{The QFI for a squeezed coherent state with probe state energy $\tilde{n}=100$. Same $\Theta$ as in Fig. \ref{okay}}
  \label{onehundred}
\end{figure}

The fidelity for the squeezed coherent state is:
\begin{equation}
\begin{split}
\mathcal{F}=&1-\frac{t^2(2(\Theta t)^4-2(\Theta t)^2+1)}{( 1-t^2 \Theta^2) (2 \Theta^2 t^2 (1-t^2 \Theta^2) +( \sinh^{-2}(r)))} d\Theta^2\\
&-|\alpha_0 t|^2 (\frac{\cos^2 (\theta)}{V_1^+}+\frac{\sin^2 (\theta)}{V_1^-}) d \Theta^2 
\end{split}
\label{fidelity}
\end{equation}

One can show that, the optimal angle is $\arg (\alpha)=\theta=0$ \cite{adaptive}. The Quantum Fisher information is simply:

\begin{equation}
\begin{split}
H(\Theta)=&\frac{4t^2(2(\Theta t)^4-2(\Theta t)^2+1)}{( 1-t^2 \Theta^2) (2 \Theta^2 t^2 (1-t^2 \Theta^2) +( \sinh^{-2}(r)))} \\
&+  \frac{4|\alpha_0 t|^2}{\Theta^2 t^2 (e^{-2 r}-1)+1}
\end{split}
\label{scqfi}
\end{equation}

We restrict the pure Gaussian probe state to finite energy with mean photon number $\tilde{n}=\sinh^2 r + |\alpha|^2$ and we optimize $H(\Theta)$ over the squeezed fraction $y=\sinh^2 r/\tilde{n}$. We note that the squeezing parameter $r$ and the amplitude $\alpha$ are the only relevant parameters. 

An estimate for $\Theta$ is needed because $H(\Theta)$ explicitly depends on the parameter we are estimating. We will graphically report the fraction of squeezed photons for maximum information if $\Theta$ is near unity. For the ideal case where $t=1$, the Quantum Fisher Information improves with more squeezing for a low number of photons as seen in figure \ref{fig2}. For $\tilde{n} \ge 100$, a large fraction of squeezed photons becomes less advantageous and it would be inefficient to squeeze all photons. 

For less than ideal transmission, there is a tendency for the maximum QFI to occur for small fractions of squeezing. In figures \ref{onen}, \ref{okay} and \ref{onehundred}, the Quantum Fisher Information is shown for three mean photon numbers $\tilde{n}=1$, $10$ and $100$ and transmission coefficient $t$ ranging from $t=0$ to $t=1.00$. For a small number of photons $\tilde{n}=1$ to $\tilde{n}=10$ and low transmission, the squeezing does not affect the information. For high transmission, the maximum information occurs for a large fraction of squeezed photons. However, for increasing number of photons $\tilde{n} \ge 100$ (see Fig. \ref{onehundred}), this maximum is attained for a smaller fraction of squeezing until squeezing becomes irrelevant and fully coherent photons are the most advantageous for all levels of loss. Furthermore, for very lossy transmission, we observed that the maximum QFI occurs for a finite fraction of squeezed photons. Nonetheless, fully coherent probe states differ negligibly from the maximum QFI of these lossy channels.

Now that we've fully characterized the QFI of a lossy quantum channel, we can apply the results to the measurement of space-time parameters.


\section{Estimating space-time parameters}
In this section, we will use the previously outlined model to estimate the space-time curvature using a lossy quantum channel. The probe state sent by Alice from Earth's surface will experience attenuation both due to scattering by the atmosphere but also from diffraction of the beam as it propagates to Bob. We will begin by presenting an approximate model for wave packets propagating in Earth's space-time as derived in Ref \cite{BRU14}. 

Earth's space-time can be approximated to be a non-rotating spherical body in the (1 + 1)- dimensional Schwarzschild metric if we assume Bob is geostationary. Therefore, the angular momentum is negligible because Alice and Bob are radially aligned. Disregarding the angular coordinates, the reduced Schwarzschild line element is $ds^2=g_{\mu \nu}dx^{\mu} dx^{\nu}=-f(r) dt^2 + \frac{1}{f(r)} dr^2$ where $f(r)=1-r_s/r$ and $r_s=\frac{2GM}{c^2}$ is Earth's Schwarzschild radius \cite{grav, wald}. An observer at radius $r=r_0$ in this metric will measure the proper time $\tau=\int ds= \sqrt{f(r_0)} t$ where $t$ is the proper time as measured by an observer at infinite distance $r=\infty$.

The electromagnetic field of a photon can be described by a bosonic massless scalar field and Klein-Gordon equation $\Box \Phi =0$ where the d'Alambertian is given by $\frac{1}{\sqrt{-g}} \partial_{\mu} \sqrt{-g}\partial^{\mu}$ \cite{modes}. Using the Eddington-Finkelstein coordinates as done in Ref \cite{BRU14}, the solutions of this equation are given by outgoing $u=ct-(r+r_s \log {|\frac{r}{r_s}-1|})$ and $v=ct+(r+r_s \log {|\frac{r}{r_s}-1|})$ ingoing waves that follow geodesics \cite{grav, wald}. It it straightforward to show that the field operator $\Phi$ is expressed as the combination of bosonic annihilation and creation operators of these outgoing and ingoing waves:

\begin{equation}
\int^{+\infty}_0 {d\omega} [\phi^{(u)}_{\omega} a_{\omega}+\phi^{(v)}_{\omega} b_{\omega}+H.c.]
\end{equation}  

Where $u$ and $v$ are the geodesic coordinates of the outgoing and incoming waves. The annihilation/creation operators obey the relations $[a_{\omega},a^{\dagger}_{\omega'}]=[b_{\omega},b^{\dagger}_{\omega'}]=\delta(\omega-\omega')$. We define localized annihilation and creation operators in terms of the frequency distribution $F(\omega_A)$ of the mode $\tilde{a}^\dagger_{\omega_A}=\int_0^{+\infty} d\omega_A F(\omega_A) e^{-i \omega_A \frac{u_A}{c}} a^{\dagger}_{\omega_A} $. 

In the Schwarzschild background, the mode Bob will receive is transformed to $\tilde{a}^\dagger_{\omega_B}=\int_0^{+\infty} d\omega_B F(\omega_B) e^{-i \omega_B \frac{u_B}{c}} a^{\dagger}_{\omega_B} $ and if Bob tunes his detector to receive Alice's frequency distribution $F(\omega_A)$, then the field can be divided into a part which matches Bob's detector and a part which does not \cite{rhode}. This is formally equivalent to a beamsplitter with transmission parameter $\Theta$. To implement this scheme, Bob would have to employ a mode selective beamsplitter transformation that extracts the desired mode \cite{ECK11}. In Appendix A, we outline a method for a mode splitter using linear optics such that the commutation relation $[a, a'^{\dagger}]=\Theta$ holds, and Bob effectively implements a beamsplitter with transmission $\Theta$.      



The frequency that Bob measures $\omega_B= \sqrt{\frac{f(r_A)}{f(r_B)}} \omega_A$ is said to be gravitationally redshifted, a famous result of general relativity \cite{wald}. For any arbitrary frequency distribution, the relation between Alice's and Bob's modes can be used to find the relation in the different reference frames \cite{qftc},

\begin{equation}
F^{(B)}_{\omega_{B,0}}(\omega_B)=(\frac{f(r_B)}{f(r_A)})^{1/4} F^{(A)}_{\omega_{A,0}} (\sqrt{\frac{f(r_B)}{f(r_A)}} \omega_B)
\label{freq}
\end{equation}

Thus the effective beamsplitter ratio is characterised by the overlap of the frequency distributions sent by Alice and received by Bob. Or equivalently, the commutation relation of the annihilation and creation operators is no longer normalized. 

\begin{equation}
[\tilde{a}_{\omega_A}, \tilde{a}^{\dagger}_{\omega_B}]=\int^{+\infty}_{0} {d \omega_B F^{*B}_{\omega_{B,0}} (\omega_B)F^{A}_{\omega_{A,0}} (\omega_B)} e^{ i\omega_B \frac{u_B-u_A}{c}}=\Theta
\label{thet}
\end{equation}
 

Where $u_B-u_A =c\tau-(r_B-r_A)-r_s \log{|\frac{r_B-r_s}{r_A-r_s}|}$ and $\tau$ is the proper time interval between Alice and Bob as measured by Bob at height $r_B$. We assume Alice and Bob directly measure their separation. Thus, Bob can tune $\tau$ such that $u_B-u_A=0$.


Since the source is not monochromatic, we need a frequency distribution for the mode, we first assume it takes the form of a normalized Gaussian wavepacket $F(\omega)=\frac{e^{-\frac{(\omega-\omega_0)^2}{4\sigma^2}}}{(2\pi \sigma^2)^{1/4}}$ centred around the frequency $\omega_0$ and with a spread of $\sigma$ \cite{gauss}. We derive a general expression for the overlap between Alice's transformed wavepacket and Bob's arbitrary choice of the Gaussian shape. For the latter, we denote Bob's detector centre frequency as $b \omega_0$ and the frequency spread $c \sigma_0$. By using equations \ref{freq} and \ref{thet}, the overlap $\Theta$ for this case is given by:

\begin{equation}
\Theta=\sqrt{\frac{2c(1-\delta)}{c^2+(1-\delta)^2}} e^{- \frac{(1-\delta-b)^2 \omega^2_{B,0}}{4(c^2+(1-\delta)^2)\sigma^2}}
\end{equation}   

Where
\begin{equation}
\delta=1-\sqrt{\frac{f(r_A)}{f(r_B)}}\approx \frac{r_s}{2} \frac{L+ r_s \log{|\frac{r_B}{r_A}|}}{r_A(r_A+L+ r_s \log{|\frac{r_B}{r_A}|})} 
\end{equation}
Where $L+r_s \log{|\frac{r_B}{r_A}|}$ is the measured distance between Alice and Bob, and $L=r_B-r_A$. This approximation holds because $r_s$ of Earth is very small compared to $r_B$ and $r_A$. We can make a further approximation and disregard the height corrections due to the geodesic in Schwarzcshild space-time since these are on the order of $r_s$. Therefore we are left with:

\begin{equation}
\delta \approx \frac{r_s}{2} \frac{L}{r_A(r_A+L)} 
\end{equation} 

However, Bob can adjust the overlap artificially by changing the shape of his detector. We assume that Bob can adjust his detector parameters to be very closely matched with a deviation of $\epsilon$ such that $b=c=1-\epsilon$. Setting $u_B-u_A=0$ in Eq. \ref{thet}, the overlap becomes: 
\begin{equation}
\Theta \approx e^{-\frac{(\delta-\epsilon)^2 \omega^2_0}{8 \sigma^2}}
\end{equation}

We denote the exponent as $x=\frac{(\delta-\epsilon)^2 \omega^2_0}{8 \sigma^2}$.  





Since $\Theta$ is equivalent to the beamsplitter transmission parameter, we can use the Quantum Fisher Information found in equation \ref{scqfi} to determine the Cramer-Rao bound and in particular we can incorporate the loss of the probe beam. However, to estimate the Schwarzschild parameter $r_s$, we must determine the corresponding Quantum Fisher informations. From the definition of fidelity, this only requires the application of the chain rule such that $H(r_s)=( \frac{d\Theta}{d r_s} )^2 H(\Theta)$. 

\section{Estimating the Schwarzschild radius with a lossy quantum probe}
In this section we will optimize our choice for the parameter $\Theta$, and consequently the probe state energy, to provide the most precise bound on the relative error $\braket{\Delta r_s}/r_s$. 

In transforming $H(\Theta)$ to $H(r_s)$, the chain rule $d\Theta=\frac{d\Theta}{dr_s} dr_s = -(\frac{ (\delta-\epsilon) \omega^2_0 \delta}{ 4 \sigma^2 r_s}) e^{-x} dr_s =-(\frac{2 x \delta}{r_s (\delta-\epsilon)})e^{-x} dr_s$ was used. 

Therefore, $H(r_s)=\frac{4 x^2\delta^2}{r_s ^2(\delta-\epsilon)^2} H(\Theta)$ for the Gaussian frequency profile. The bound for the relative error in the Schwarzschild radius is

\begin{equation}
 \frac{\Delta r_s}{r_s}\ge \frac{1}{ r_s \frac{d\Theta}{d r_s} \sqrt{N H(\Theta)}} = \frac{4 \sigma^2 }{(\delta-\epsilon) \omega^2_0 \delta e^{-x}\sqrt{N H(\Theta)}}
 \label{rsbound}
 \end{equation}

\begin{figure}
\includegraphics[width=1.12 \linewidth]{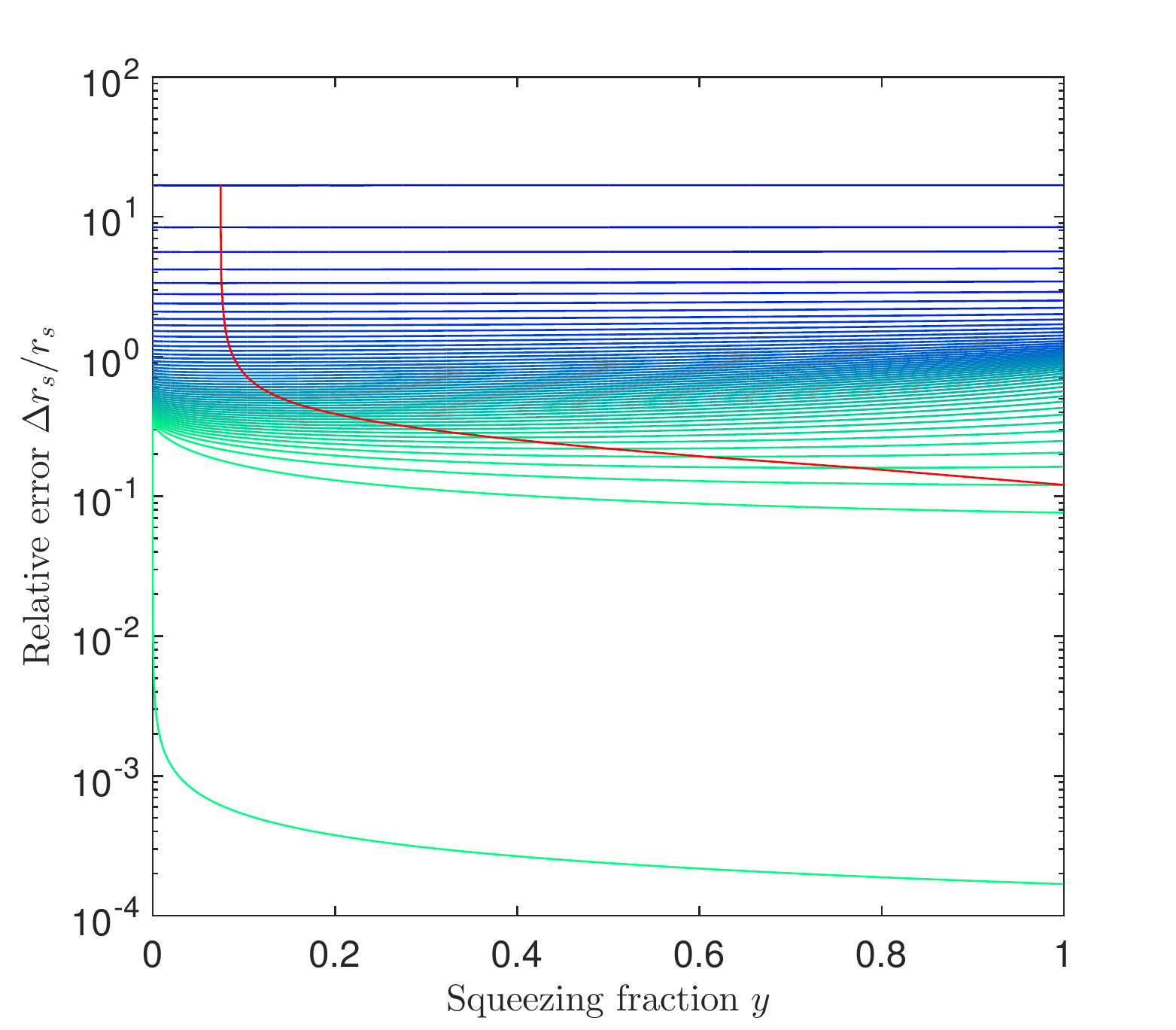}
\caption{Relative error in the Schwarzschild radius versus squeezing fraction at the operating point $(\delta-\epsilon) \omega=1$ Hz ($x=3 \times 10^{-10}$) with $\tilde{n}=2$  from $t=0.02$ (blue) to $t=1.00$ (green) in $0.02$ intervals. The number of measurements are $N=\sigma/10=2 \times 10^2$. We take $r_A=6.37 \times 10^6$m, $r_B=42.0\times 10^6$ m, $\sigma=2000$, $\omega_0=700$ THz and hence $\delta=6.0 \times 10^{-10}$}
\label{pulse1}
\end{figure}

\begin{figure}
\includegraphics[width=1.12 \linewidth]{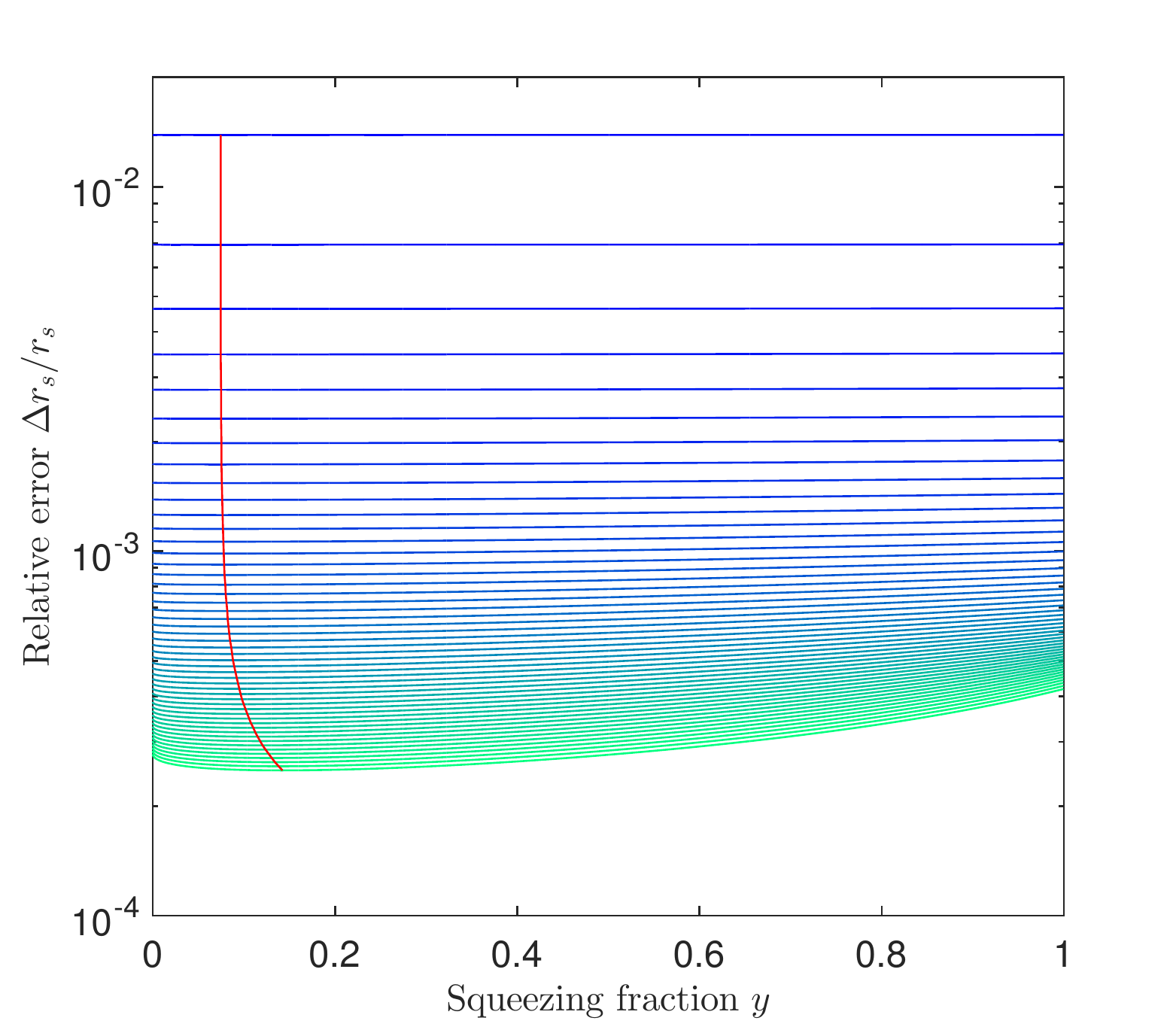}
\caption{$(\delta-\epsilon_{opt}) \omega=2\sigma $ Hz ($x_{opt}=1/2$) with $\tilde{n}=2$. This is the optimal operating point for coherent states. The parameters are the same as Fig. \ref{pulse1}}
\label{pulse2}
\end{figure}

\begin{figure}
\includegraphics[width=1.12 \linewidth]{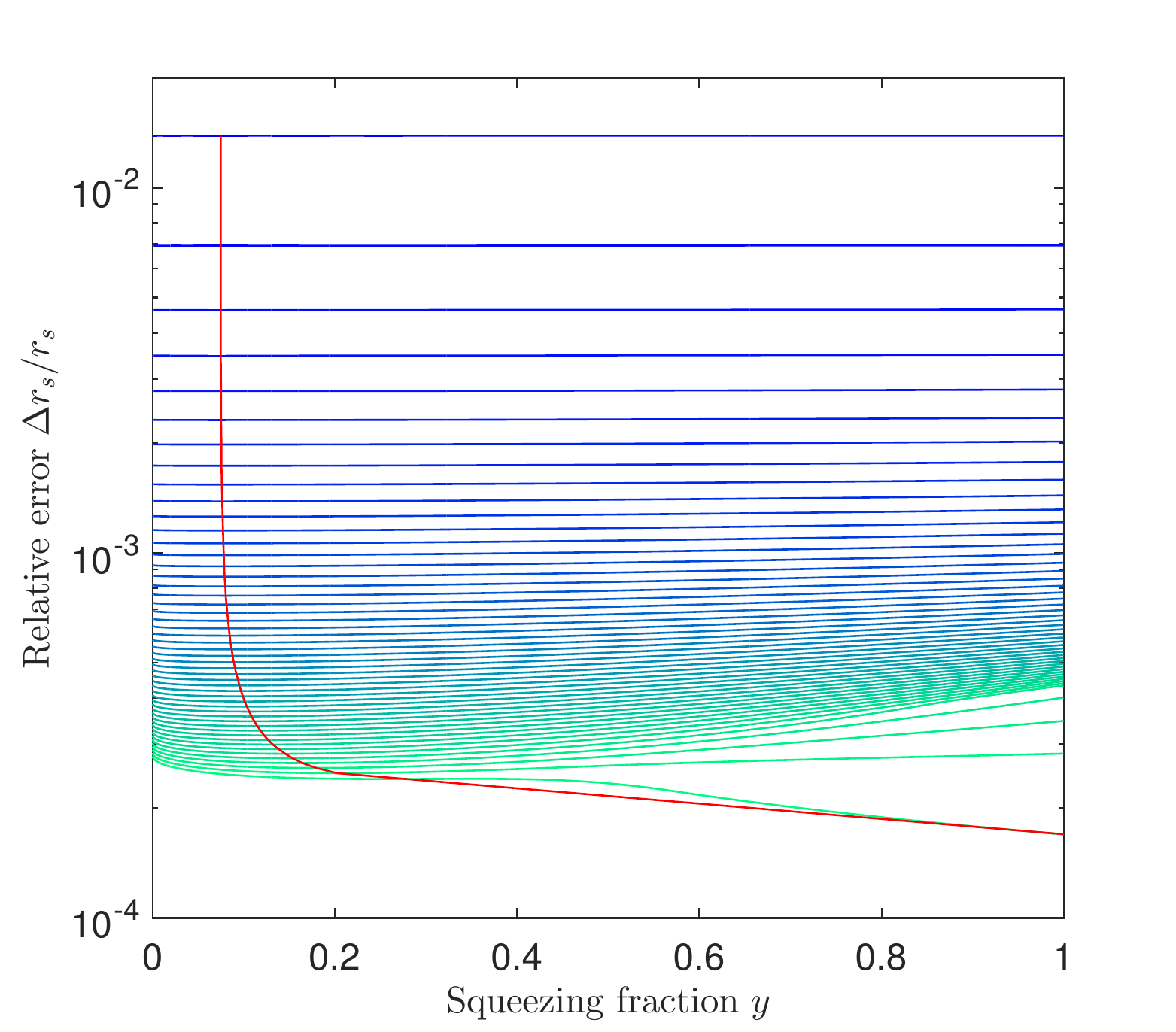}
\caption{For each squeezing fraction and transmission, we've optimized for the best $\epsilon$. Every other parameter is the same as Fig. \ref{pulse1}}
\label{optgauss}
\end{figure}
 
Let us explore some properties of this equation. From Eq. \ref{rsbound} and the Quantum Fisher information of a coherent state $H(\Theta)=|t\alpha|^2$, we can see that as $t \rightarrow 1$ and $\Theta \rightarrow 1$ which corresponds to $\epsilon \rightarrow \delta+\frac{\Delta^2}{4 \delta \Omega_0^2} $, the variance diverges and it is impossible to measure the Schwarzschild radius $r_s$ with coherent states around the point $\Theta=1$. However, if the probe energy is completely squeezed, then the denominator reduces to:

\begin{equation}
\begin{split}
&\lim_{\epsilon \rightarrow \delta,  t \rightarrow 1}{(\delta-\epsilon) \omega^2_0 \delta e^{-x} \sqrt{N H(\Theta)}}\\
&= (\delta-\epsilon) \omega^2_0 \delta e^{-x} \sqrt{\frac{4 N \sinh^2 r}{1-\Theta^2}}
\end{split}
\end{equation}
 
We can make further approximations about $\Theta^2 \approx 1- 2x$. Thus, since $\frac{1}{\sqrt{2 x}} \propto \frac{1}{\delta-\epsilon}$ this cancels out, and the limit is well behaved. The relative error bound of $\frac{\Delta r_s}{r_s}$ approaches:

\begin{equation}
 \frac{\Delta r_s}{r_s}\ge \frac{\sigma}{\omega_0 \delta \sqrt{N \sinh^2 r} }
 \label{thet1squeeze}
\end{equation}

It is evident that coherent states do not contribute to the error bound if $\Theta=1$ and Bob matches up his wavepacket with the one he receives $\delta=\epsilon$. Squeezing is absolutely necessary to detect the small deviation from $\Theta=1$. As seen in Fig. \ref{pulse1}, if the overlap is $\Theta=1-3 \times10^{-10}$, the error of $r_s$ is far too large for lossy channels. The only exception is when $t=1$ which depends strongly on the amount of squeezing. 

However, if Bob's detector is using Alice's original wavepacket $\epsilon=0$ then the error bound for coherent states is sensible:

\begin{equation}
 \frac{\Delta r_s}{r_s}\ge \frac{2 \sigma^2}{ \omega^2_0 \delta^2 e^{-x} \sqrt{N |\alpha|^2} }
\end{equation}

And no squeezing is necessary. Furthermore, we determine the optimal $\epsilon$ for which the coherent state is most advantageous. We simply minimize \ref{rsbound} with respect to $\epsilon$ to obtain $\epsilon_{opt}=\delta-\frac{2 \sigma}{\omega_0}$ and $x_{opt}=\frac{1}{2}$.  The $r_s$ lower error bound for $t=1$ is thus:

\begin{equation}
 \frac{\Delta r_s}{r_s}\ge \frac{3.3 \sigma }{\omega_0 \delta \sqrt{N |\alpha|^2}}
 \label{rsboundb}
 \end{equation}

As seen in Fig. \ref{pulse2}, the lossy channels now have reasonable error bounds. Finally, we can determine the optimal $\epsilon$ for a given squeezing fraction and transmission as seen in Fig. \ref{optgauss}. In figure \ref{optgauss}, $\frac{\Delta r_s}{r_s} $ is plotted against the fraction of squeezed photons for $\tilde{n}=2$ average photons of the initial probe state.   
For low transmission $t$, the minimum error occurs for a small fraction of squeezing that is less than $10 \%$. Therefore, it is more advantageous to use coherent photons for heavily attenuated signals. However, for almost perfect transmission, it is considerably more advantageous to squeeze $100 \%$ of the photons. 

In figures 6 to 15 we adopt parameters similar to Ref. \cite{BRU14}. We assume that Alice is on the surface of Earth $r_A=6.37 \times 10^6$m and Bob is in geostationary orbit $r_B=42.0\times 10^6$ m. With these parameters, we obtain $\delta=6.0 \times 10^{-10}$. In contrast to Ref. \cite{BRU14}, we relate the number of measurements $N$ per second to the frequency width $\sigma=2$ kHz. We assume that the length of the pulse is $\Delta t = 1/\sigma= 0.5$ $ms$ and thus within one second, we can have up to $10^3$ measurements. For minimal correlation between pulses, we assume that one pulse has $50$ milliseconds of space, and as a rule of thumb $N=\sigma/10=2\times 10^2$ measurements. Counterintuitively, by making the width of $\sigma$ smaller and thus the number of measurements smaller, we find $r_s$ is more precise. 
 




\subsection{Rectangular frequency profile}
It is apparent that the overlap between Alice's and Bob's Gaussian wavepackets is very near unity. However, it is known that Gaussian wave-packets are the best at maintaining their overlap given a displacement \cite{opt}. This means they are least optimal for our purposes. We require a frequency profile that is more sensitive. Hence, we consider a rectangular frequency profile. In the time-domain, this would correspond to a sinc$(kt)$ function. 

For a normalised rectangular function of width $\sigma_0$ and height $1/\sqrt{\sigma_0}$ at frequency $\omega_0$, we wish to calculate the overlap with the transformed rectangular function in Bob's reference frame. From equation \ref{freq}, Bob measures the width $\sqrt{\frac{f(r_A)}{f(r_B)}} \sigma_0$ and height $\frac{1}{\sqrt{\sigma_1}}=\frac{1}{\sqrt{\sqrt{\frac{f(r_A)}{f(r_B)}}\sigma_0}}$ centred at frequency $\omega_1=\sqrt{\frac{f(r_A)}{f(r_B)}} \omega_0$. The rectangular function profile must be normalised $\int{|F_B(\omega_B)|^2}=1$. Since $\sqrt{\frac{f(r_A)}{f(r_B)}} < 1$, the transformed rectangular function will have lower frequency and smaller overall width but larger height. Thus, making use of equation \ref{thet}, the overlap is:

\begin{equation}
\Theta=\begin{cases}
 [\omega_B+\frac{\sigma_B}{2}-(\omega_A-\frac{\sigma_A}{2})]\frac{1}{\sqrt{\sigma_B \sigma_A} }, & \text{$\omega_B+\frac{\sigma_B}{2}> \omega_A -\frac{\sigma_A}{2}$} \\
 0, & \text{$\omega_B+\frac{\sigma_B}{2}< \omega_A -\frac{\sigma_A}{2}$}
 \end{cases}
\end{equation}

Bob can adjust his detector by varying the central frequency $\omega_0$ by an arbitrary factor $b$ and also the frequency spread $\sigma_0$ by a factor of $c$. 
Therefore,

\begin{equation}
\begin{split}
\Theta&=\frac{[(\sqrt{\frac{f(r_A)}{f(r_B)}}  \omega_0+\frac{\sqrt{\frac{f(r_A)}{f(r_B)}} \sigma_0}{2})-(b \omega_0-\frac{c \sigma_0}{2})]}{(\frac{f(r_A)}{f(r_B)})^{1/4} \sqrt{c} \sigma_0}\\
&=\frac{\omega_0}{\sigma_0}\left(\frac{1}{\sqrt{c}}\left(\frac{f(r_A)}{f(r_B)}\right)^{1/4}-\frac{b}{\sqrt{c}}\left(\frac{f(r_A)}{f(r_B)}\right)^{-1/4}\right) \\
&+\frac{1}{2}\left(\frac{1}{\sqrt{c}}\left(\frac{f(r_A)}{f(r_B)}\right)^{1/4}+\sqrt{c} \left(\frac{f(r_A)}{f(r_B)}\right)^{-1/4}\right)
\end{split}
\end{equation}

We note the useful approximation $\left(\frac{f(r_A)}{f(r_B)}\right)^{1/4} \approx 1-\frac{r_s L}{4r_A r_B}=1-\frac{\delta}{2}$ and $\left(\frac{f(r_A)}{f(r_B)}\right)^{-1/4} \approx 1+\frac{r_s L}{4r_A r_B}=1+\frac{\delta}{2}$. This approximation holds if $r_s$ is extremely small as is the case for Earth. We can further set the factors $b=c=1-\epsilon$. We make the necessary approximations and keep terms to first order:

\begin{equation}
\begin{split}
\Theta&=\frac{\omega_0}{\sigma_0}  \left( (1+\frac{\epsilon}{2}) (1-\frac{\delta}{2}) -(1-\frac{\epsilon}{2})(1+\frac{\delta}{2})\right) \\
&+ \frac{1}{2} \left( (1+\frac{\epsilon}{2}) (1-\frac{\delta}{2}) +(1-\frac{\epsilon}{2})(1+\frac{\delta}{2})\right)\\
&=1 - \frac{\omega_0| \delta-\epsilon|}{\sigma_0} 
\end{split}
\label{rec}
\end{equation}

In the last step, we've generalised to the case when Bob adjusts his detector to overestimate the frequency spread and central frequency. Thus $\epsilon> \delta$ but the overlap will remain smaller than unity, as expected.
The expression above is only valid if $\omega_0>> \sigma_0$ because we are disregarding the negative frequencies. For sufficiently large $\frac{\omega_0}{\sigma_0}$, the rectangular frequency profiles will no longer overlap because of the gravitational redshift $\Theta=0$. The modes remain completely orthogonal for $\frac{\omega_0}{\sigma_0} \ge \frac{1}{|\delta-\epsilon|}$. The error lower bound for $r_s$ is:

 \begin{equation}
 \frac{\Delta r_s}{r_s}\ge \frac{\sigma_0}{\omega_0 \delta \sqrt{N H(\Theta)}}
 \label{rsgauss}
 \end{equation}
We note that there is a discontinuity at $\delta=\epsilon$ and the derivative w.r.t. $r_s$ of the absolute value is undefined at this point. However, in reality we adopt a continuous frequency distribution and thus $\Theta$ will have a well defined derivative. 
We note that for a coherent state, in the limit that $t \rightarrow 1$ and $\epsilon \rightarrow \delta $, the error lower bound for $r_s$ is 
 
 \begin{equation}
  \frac{\Delta r_s}{r_s}\ge \frac{\sigma}{2 \omega_0 \delta \sqrt{N |\alpha|^2}}
 \end{equation}
 
  \begin{figure}
\includegraphics[width=1.12 \linewidth]{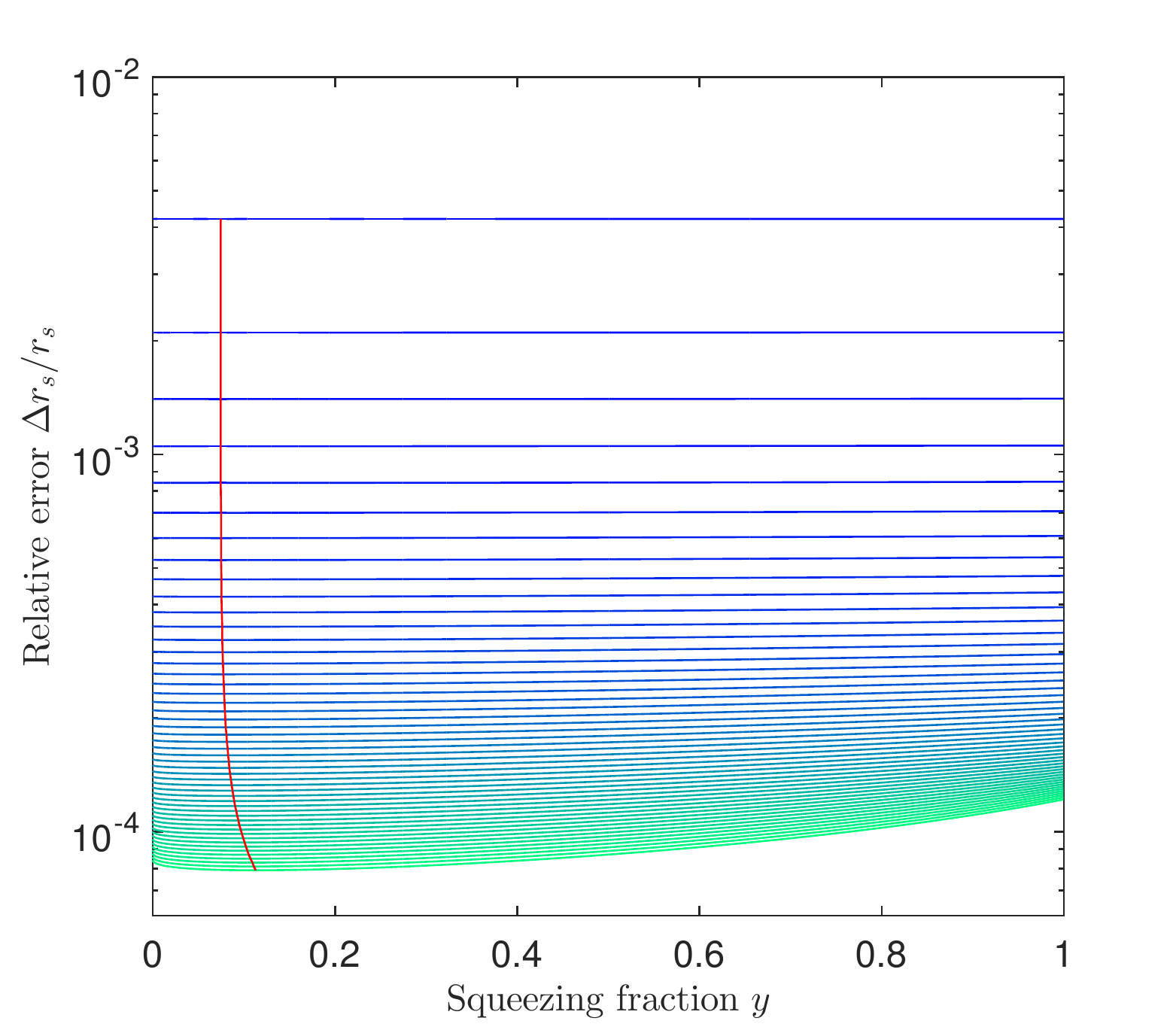}
\caption{Relative error in the Schwarzschild radius with total probe state energy $\tilde{n}=2$ for rectangular frequency profile. Squeezing is not effective. The number of measurements are $N=2\times10^2$. The transmission coefficient varies $t=0.02$ (blue) to $t=1.00$ (green) in $0.02$ intervals. (Parameters: $(\delta-\epsilon) \omega=\sigma/2$ Hz ($x=1/2$) with $\tilde{n}=2$ and $N=2 \times 10^2$ measurements) }
\label{rsquare1}
\end{figure}
 
\begin{figure}
  \includegraphics[width=1.12 \linewidth]{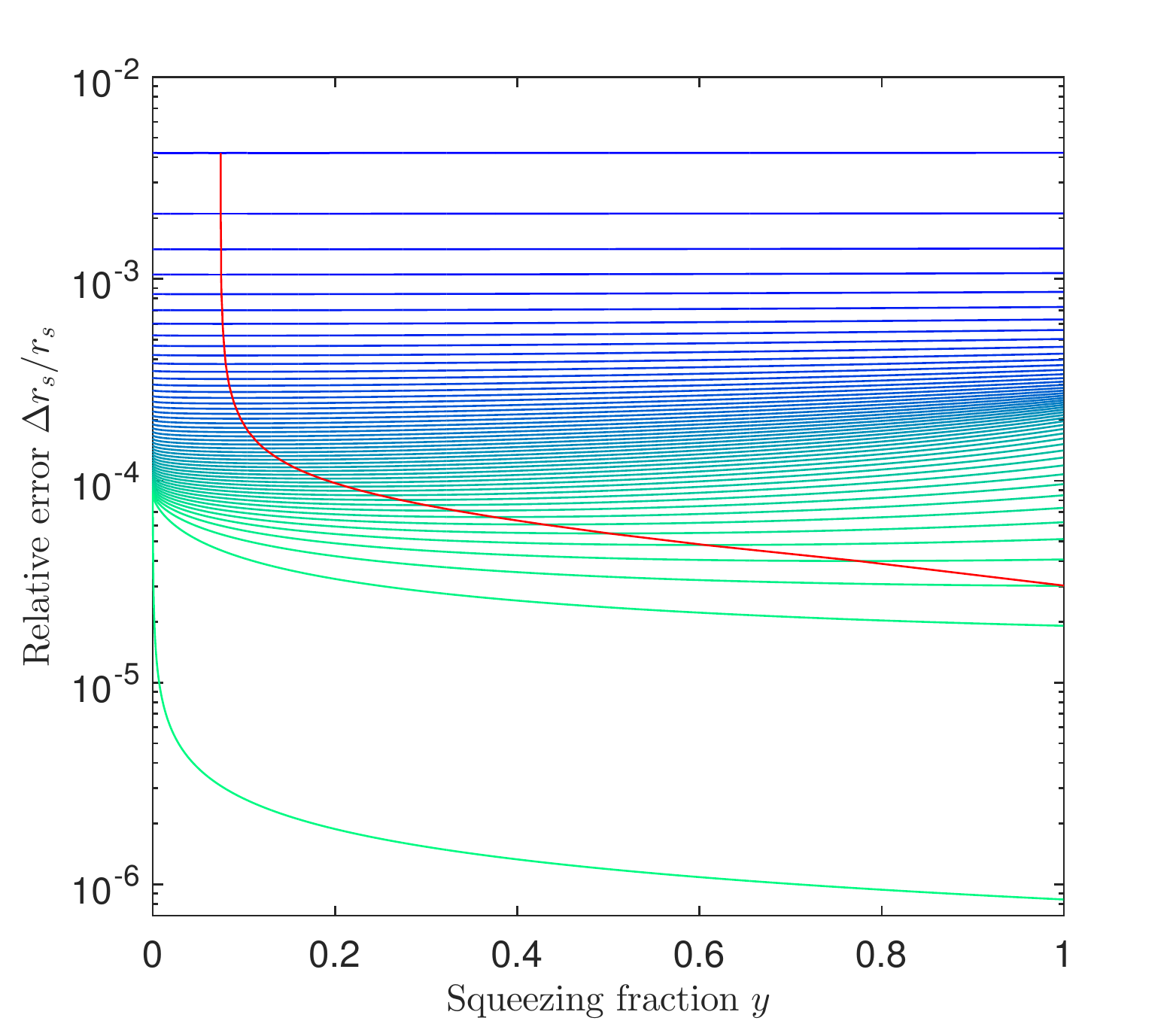}
  \caption{Same parameters as in Fig. \ref{rsquare1} with the exception $(\delta-\epsilon) \omega=1$ Hz ($x=5 \times 10^{-5}$) with $\tilde{n}=2$. In this regime, the rectangular frequency profile does $10^2$ better than the optimal Gaussian point if the initial probe state is fully squeezed}
\label{rsquare2}
\end{figure}

\begin{figure}
\includegraphics[width=1.12 \linewidth]{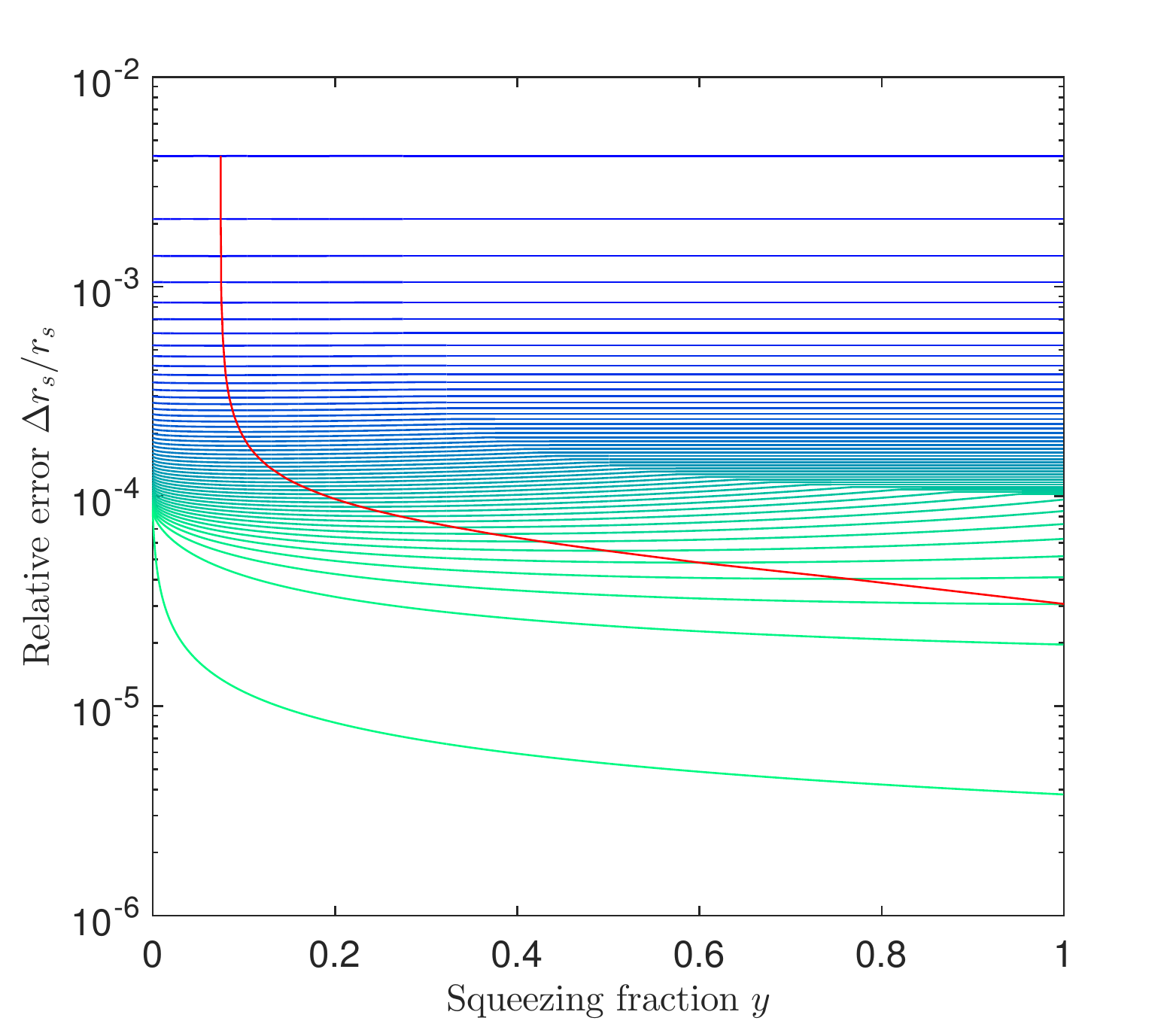}
\caption{For each squeezing fraction and transmission, we've optimized for the best $\epsilon$ (not including $t=1$ which is optimal at $\delta=\epsilon$ and approaches $0$).}
\end{figure}

This behaviour is evident in Fig. \ref{rsquare1}, where the overlap is $\Theta=1/2$, there is small dependence on the squeezing fraction. Comparatively, the optimal point of the Gaussian is up to a factor of 5 worse than for the same overlap using the rectangular frequency profile. 
Conversely, for a fully squeezed probe state, we can express the error bound as:
 
\begin{equation}
 \frac{\Delta r_s}{r_s}\ge \frac{\sqrt{\sigma_0 (\delta -\epsilon)}}{\sqrt{2 \omega_0} \delta  \sqrt{N \sinh^2 r }}
 \end{equation}
 
This indicates that the bound can approach $0$ up to an error in matching up the exact frequency distribution. For example, if Bob's detector guesses the correct distribution within $(\delta-\epsilon) \omega=1$ Hz then a perfect channel would have $10^{-6}$ precision. As seen in Fig. \ref{rsquare2}, squeezing is clearly advantageous. In comparison to the best Gaussian precision, the rectangular frequency profile does 2 orders of magnitude better.

\begin{figure}
\includegraphics[width=1.12 \linewidth]{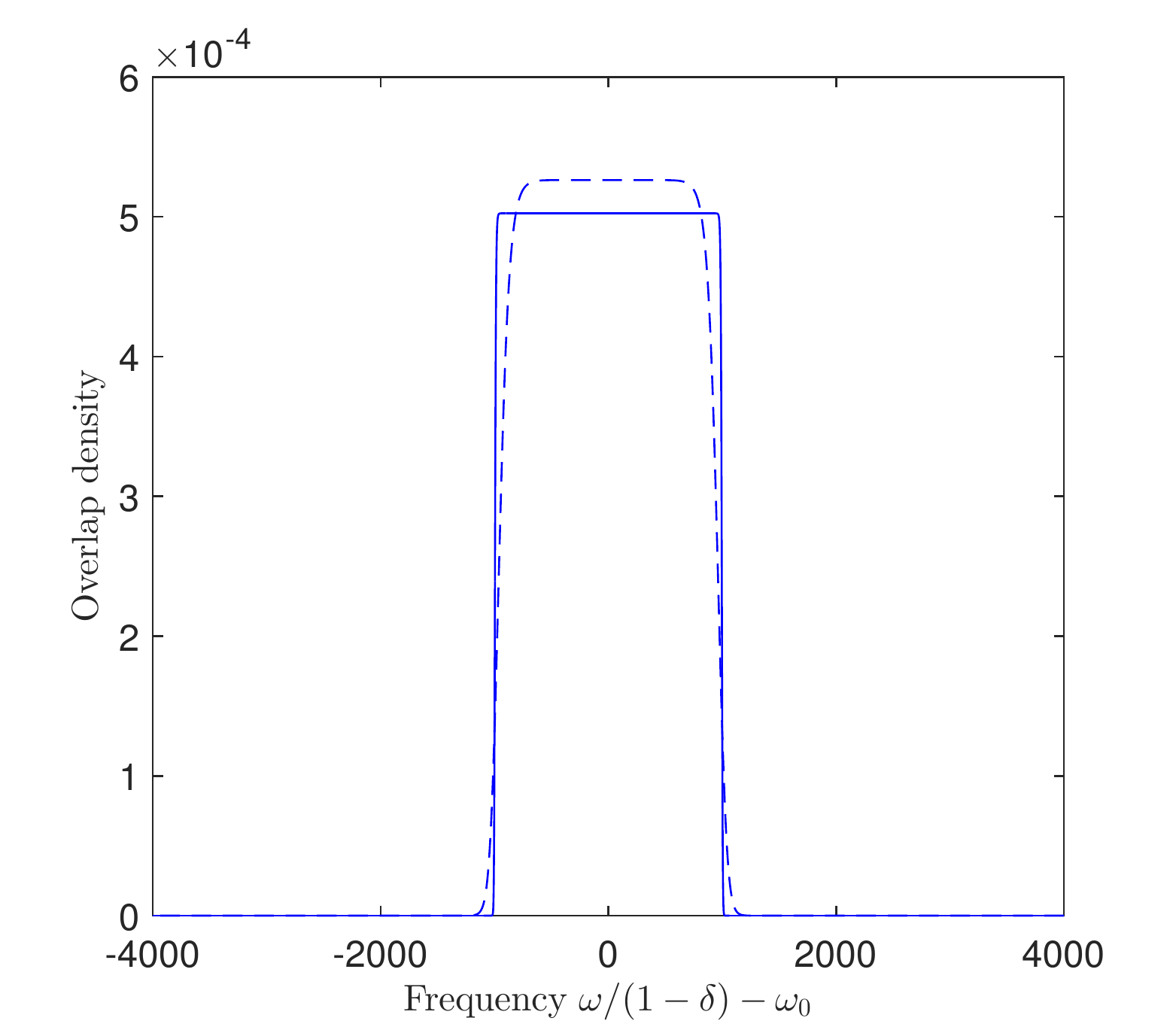}
\caption{The frequency profile squared for $\Delta=0.01$ (solid line) and $\Delta=0.1$ (dashed line) plotted against the shifted and rescaled frequency $\frac{\omega}{1-\delta_0}-\omega_0$. We take $r_A=6.37 \times 10^6$m, $r_B=42.0\times 10^6$ m, $\sigma=2000$, $\omega_0=700$ THz and hence $\delta=6.0 \times 10^{-10}$ (the same as previous section).}
\label{delta001}
\end{figure}

\begin{figure}
\includegraphics[width=1.12 \linewidth]{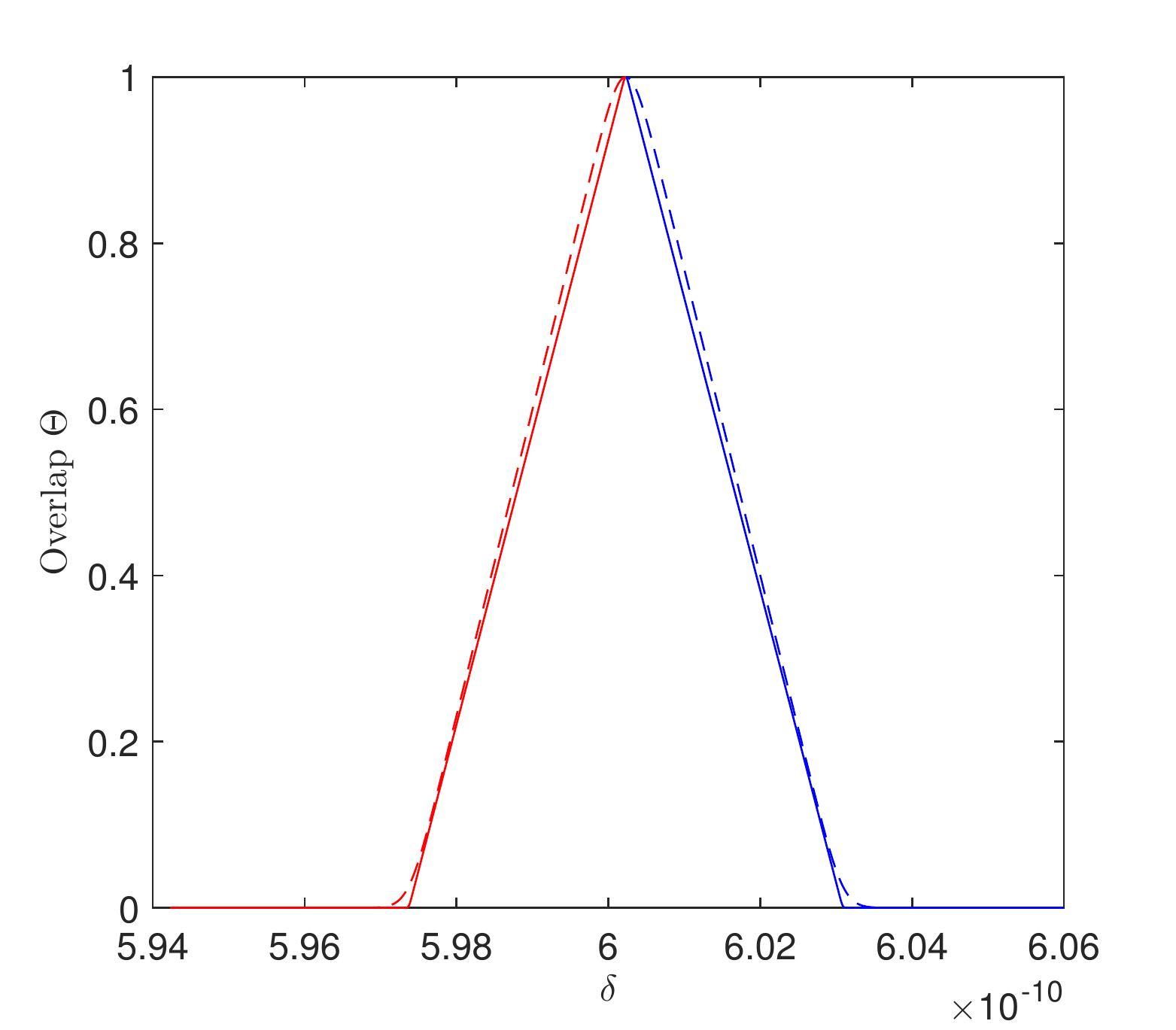}
\caption{ We vary $\delta$ around the fixed point $\epsilon=\delta_0$ and plot the overlap $\Theta$. Here we see a sharp turn at this point for $\Delta=0.01$ (solid line) but for $\Delta=0.1$ (dashed line), the $\Theta$ dependence on $\delta$ is smoother. The parameters are the same as Fig. \ref{delta001}}
\label{thetdelt}
\end{figure}

\begin{figure}
\includegraphics[width=1.12 \linewidth]{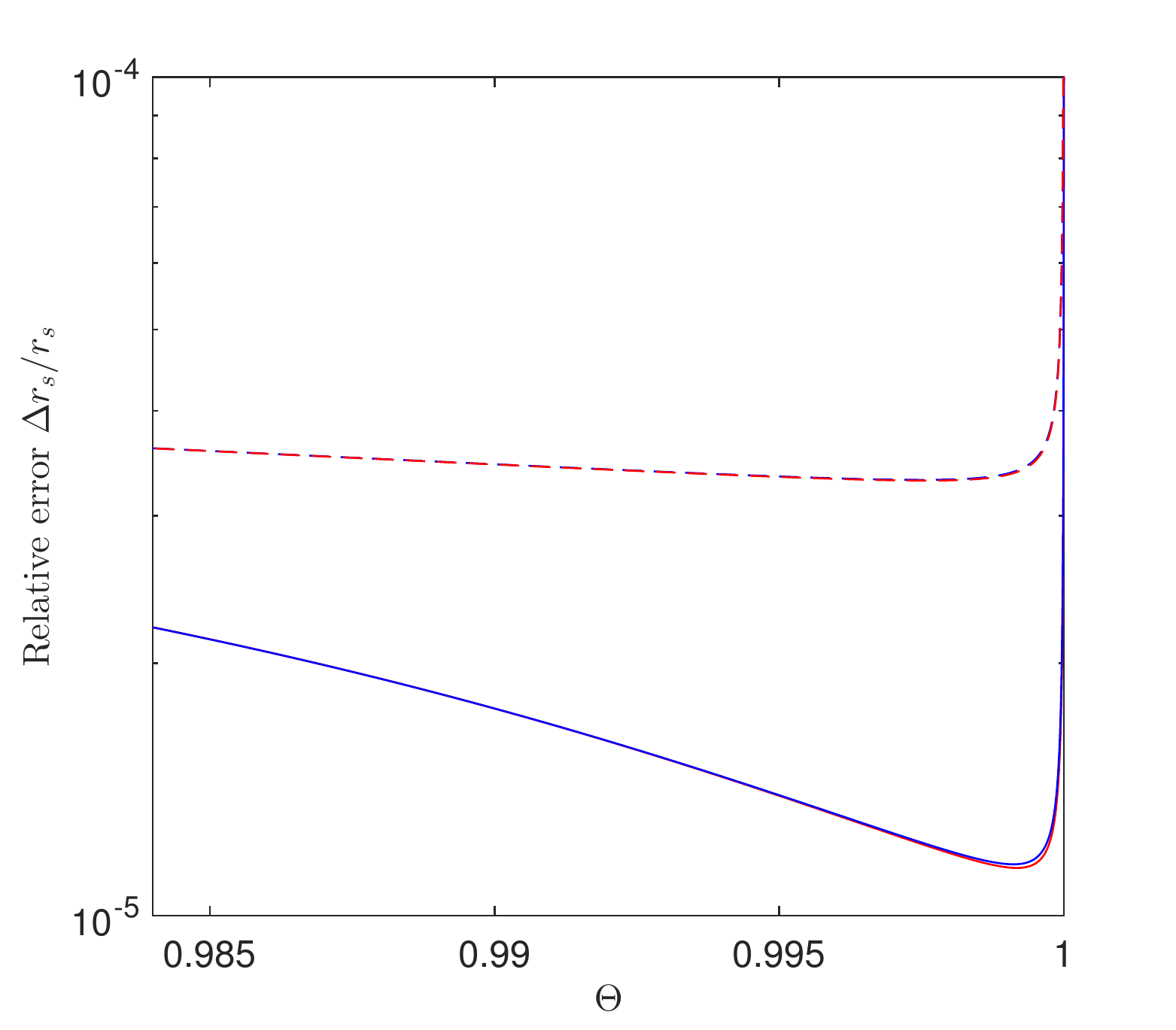}
\caption{ At $t=1$ for a fully squeezed probe state of $\tilde{n}=2$, we plot the limit as $\Theta \rightarrow 1$. We note that the optimal point is before this limit. The reason for this is the competing effect of $\frac{d \Theta}{d r_s}$. In this case, the derivative $\frac{d \Theta}{d r_s}=0$ at $\Theta=1$ and we are not impeded by the discontinuity that arose in the rectangular frequency profile. The red line is $\delta^{-}$ from below and the blue line from above $\delta^{+}$. These two lines are essentially equivalent. Note: the solid line corresponds to $\Delta=0.01$ and the dashed line to $\Delta=0.1$.}
\label{thetlim}
\end{figure}

\begin{figure}
\includegraphics[width=1.12 \linewidth]{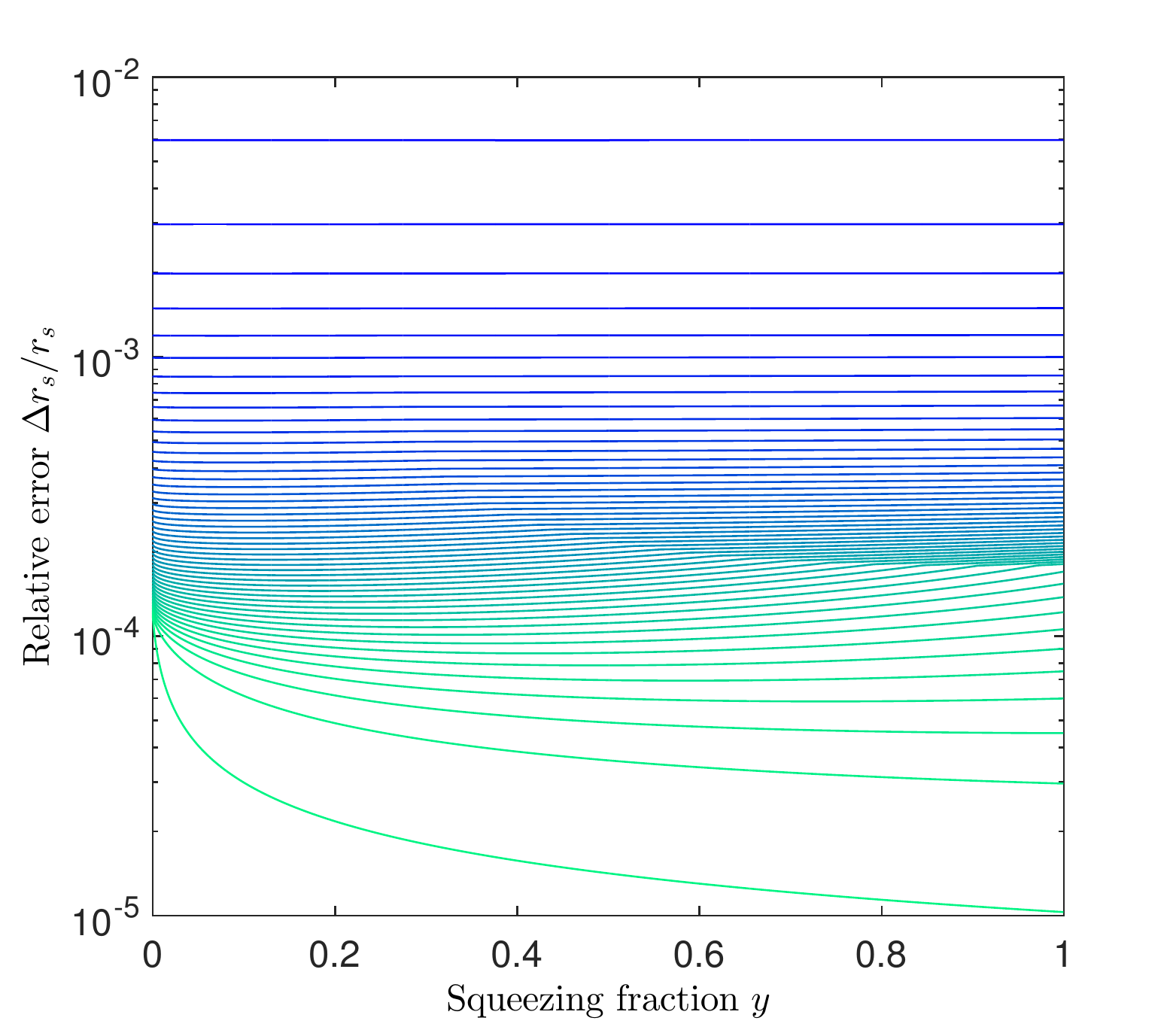}
\caption{ The Schwarzschild radius lower error bound optimized for each squeezing fraction $y$ and transmission parameter $t$ for the smoothed rectangular function.  Using the $\Theta$ in Fig. \ref{thetdelt} we determined $r_s \frac{d \Theta}{d r_s}=r_s \frac{d \delta}{d r_s} \frac{d \Theta}{d \delta}$ and determined the optimal $\Theta$ that minimizes $\Delta r_s / r_s$. We note that $t=1$ is bounded because $ \frac{d \Theta}{d r_s} \rightarrow 0$ as $\delta \rightarrow \epsilon$ and approaches a limit. (Other parameters: $\Delta=0.01$, $\tilde{n}=2$)}
\label{tanhopt}
\end{figure}

\section{More Realistic Scenarios}
\subsection{Non-ideal rectangular frequency profile}

The shape of the rectangular frequency we've used is an ideal representation with infinitely sharp edges which is unphysical. We can smooth the edges using a $\tanh {\omega}$ function as follows:

\begin{equation}
F(\omega_B)=\frac{\tanh{\frac{\sigma+2(\omega_B-\omega_0)}{\Delta \sigma}}+\tanh{\frac{\sigma-2(\omega_B-\omega_0)}{\Delta \sigma}}}{2 \sqrt{\Delta \sigma(-\frac{1}{2}+\frac{1}{\Delta}\coth{\frac{2}{\Delta}})}}
\label{tanh}
\end{equation}

As the parameter $\Delta \rightarrow 0$, the frequency profile approaches a rectangular function. In the time domain, the Fourier transform of \ref{tanh} is proportional to $\frac{\sin{\sigma t}}{\sinh{\Delta \pi \sigma t/2}}$. The function falls off exponentially depending on $\Delta$. We use the transformation in Eq. \ref{freq} and we similarly give Bob the freedom to choose the frequency spread $b \sigma_0$ and central frequency $b \omega_0$ of his detector. The overlap is thus (after normalization of Eq. \ref{tanh}):

\begin{equation}
\begin{split}
\Theta=\int d\omega & \frac{(\tanh[\frac{\sigma+\frac{\omega}{a} -\omega_0}{\Delta \sigma}]+\tanh[\frac{\sigma-\frac{\omega}{a} +\omega_0}{\Delta \sigma}])}{4\sqrt{a \Delta \sigma (-1+\frac{2 \coth[\frac{2}{\Delta}]}{\Delta})}} \\
& \times \frac{(\tanh[\frac{\sigma+\frac{\omega}{b} -\omega_0}{\Delta \sigma}]+\tanh[\frac{\sigma-\frac{\omega}{b} +\omega_0}{\Delta \sigma}])}{\sqrt{b \Delta \sigma (-1+\frac{2 \coth[\frac{2}{\Delta}]}{\Delta})}} 
\end{split}
\end{equation}

We can further simplify this equation and group the constants $\Theta=K \theta (\omega)$, where the proportionality constant is

\begin{equation}
K=\frac{\tanh[\frac{2}{\Delta}]^2}{\Delta \sigma \sqrt{a b}  (-1+\frac{2 \coth[\frac{2}{\Delta}]}{\Delta})}
\end{equation}

And 

\begin{equation}
\begin{split}
\theta (\omega)=&  \int  \frac{d\omega}{(1+\cosh [\frac{4(-a \omega_0+\omega)}{\Delta a \sigma}] / \cosh [\frac{2}{\Delta}])} \\
& \times \frac{1}{(1+\cosh [\frac{4(-b\omega_0+\omega)}{ \Delta b \omega_0}] / \cosh [\frac{2}{\Delta}])} 
\end{split}
\label{smallthet}
\end{equation}

There is no definite form of this integral. However, we can make some approximations since $a=1-\delta$ and $b=1-\epsilon$ where $\delta$ and $\epsilon$ are very small. We also can suppose that the bounds of the integral are very close to the central frequency $\omega_0$ with the bounds extending over a region of $\sigma$ across the central frequency. For the choice of $\Delta=0.1$ and $\Delta=0.01$, the overlap of the latter is approximately rectangular as seen in Fig. \ref{delta001}. Furthermore, the overlap as a function of $\delta$ is continuous at the point $\delta=\epsilon$ and the derivative with respect to $r_s$ exists in contrast to the ideal rectangular frequency profile. Approaching $\delta^{-}$ from below has the same minimum as approaching from above (as seen in Fig. \ref{thetlim}) since the frequency profile is symmetric. For the larger $\Delta=0.1$, the same behaviour occurs but the relative error of $r_s$ is larger than for the case of $\Delta=0.01$.  

In Fig. \ref{tanhopt}, we present the Schwarzschild radius lower error bound optimized for each squeezing fraction $y$ and transmission parameter $t$.  Using the $\Theta$ in Fig. \ref{thetdelt} we determined $r_s \frac{d \Theta}{d r_s}=r_s \frac{d \delta}{d r_s} \frac{d \Theta}{d \delta}$ and thus the optimal $\Theta$ that minimizes $\Delta r_s / r_s$. The behaviour is very similar to the rectangular frequency profile. We note that $t=1$ is bounded because $ \frac{d \Theta}{d r_s} \rightarrow 0$ as $\delta^{\pm} \rightarrow \epsilon$ and approaches a limit. 

\subsection{Large coherent pulses with small squeezing}
For large scale interferometers such as GEO and the Laser Interferometer Gravitational-Wave Observatory (LIGO), it has been shown  that large coherent sources of light with additional squeezing can significantly improve the detector sensitivity \cite{ligo}. 

\begin{figure}
  \includegraphics[width=1.12 \linewidth]{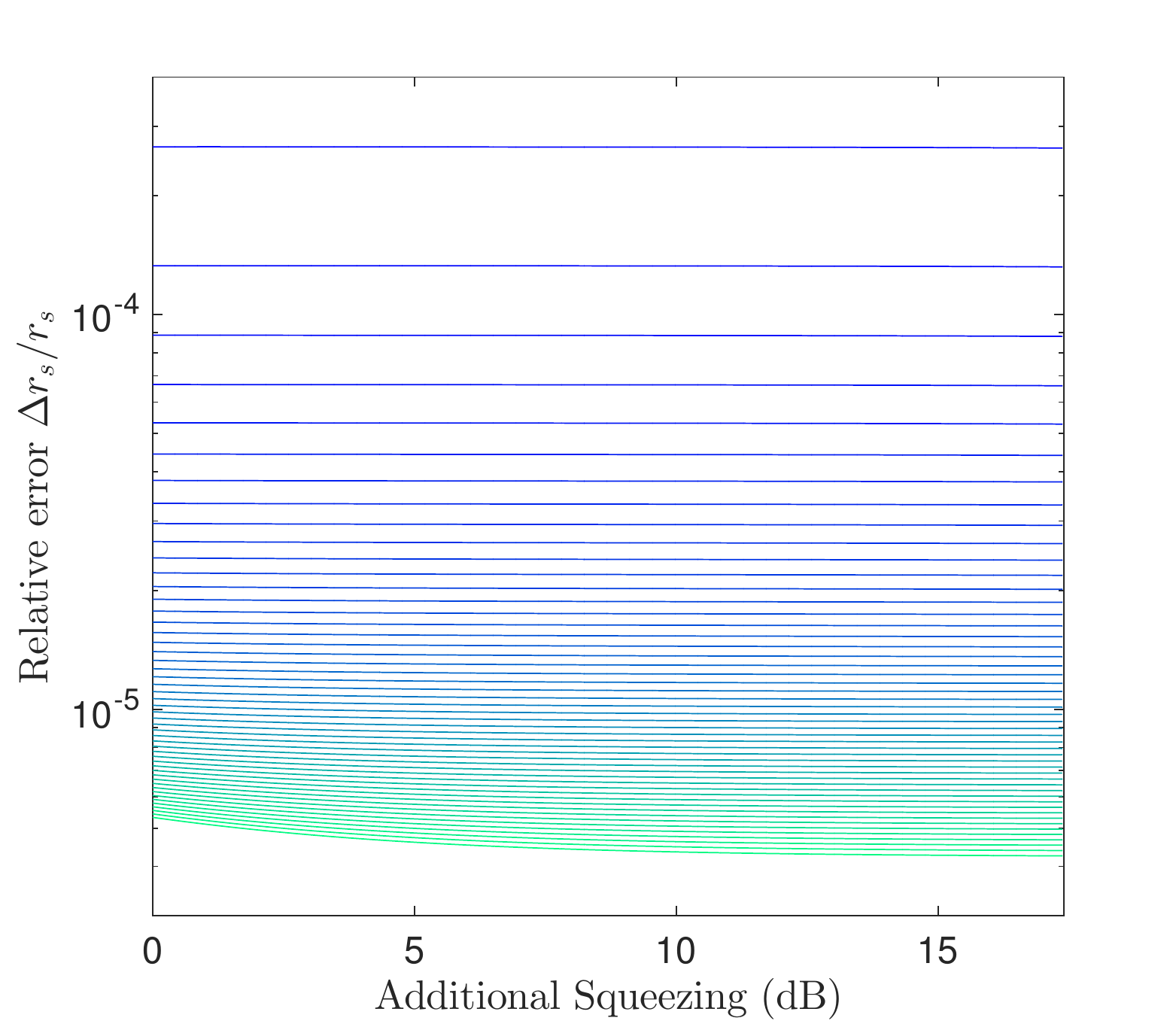}
  \label{fig:sub2}
\caption{The relative Schwarzschild error for a coherent pulse of energy $\tilde{n}=1000$ with injection of additional squeezing for Gaussian frequency profile at optimal operating point $x_{opt}=1/2$. The units of squeezing is expressed in decibels with respect to the shot noise quadrature variance. Squeezing has little effect on the lossy channels. (Parameters: $(\delta-\epsilon_{opt})\omega_0=2 \sigma $ Hz and all other parameters the same as Fig. \ref{delta001})}
\label{add100}
\end{figure}

For the Gaussian frequency profile, operating at the optimal point $x_{opt}=1/2$ requires large coherent states rather than squeezed states. As seen in Fig. \ref{add100}, the addition of squeezing does not significantly improve the precision.  

\begin{figure}
\includegraphics[width=1.12 \linewidth]{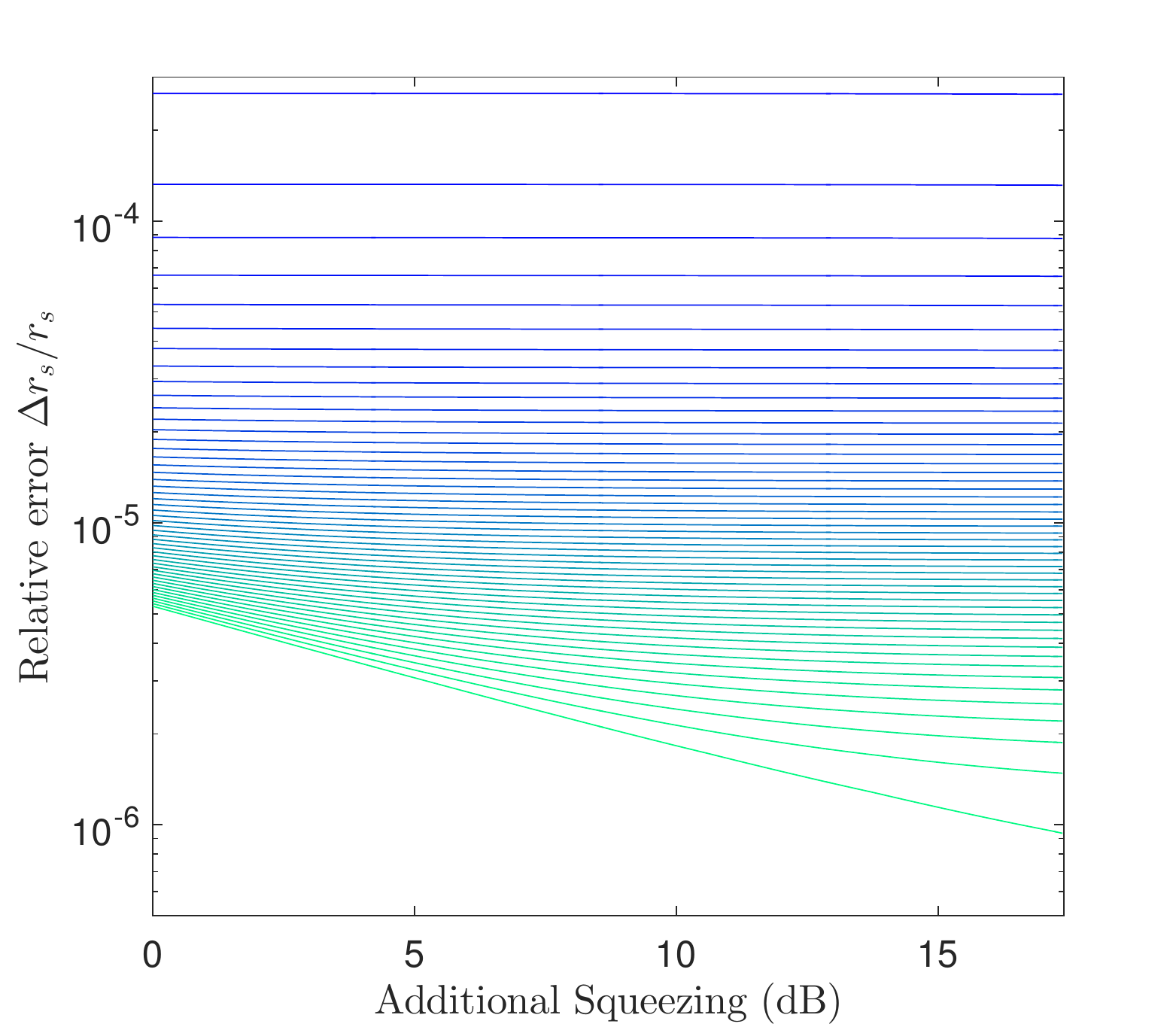}
\caption{ As in Fig. \ref{add100} but for $\Delta=0.01$ rectangular frequency profile at the optimal operating point $\Theta=0.999$ with $\tilde{n}=1000$. For almost no loss, squeezing improves the error bound up to a factor of $5 \times 10^{-1}$ for $17.4$ dB of squeezing. }
\label{add1000}
\end{figure}

Conversely, for the rectangular frequency distribution, squeezing significantly improves the precision. As seen in Fig. \ref{add1000}, for good transmission coefficients, squeezing increases the precision up to a factor $10$ for $17.4$ dB of squeezing. Current state-of-art technologies have been able to achieve squeezing of up to $12.7$ dB \cite{eb}. In both cases, the error improves with additional coherent photons.

\subsection{Optimal position of Bob}
Up to this point we have assumed that Bob is located in a geo-stationary orbit and that different levels of loss can be achieved. We consider a more realistic scenario in which the loss is a function of how high Bob is and investigate the position of Bob for which the relative Schwarzschild radius error is minimal. We model the attenuation with distance of the Gaussian beam using the characteristic Rayleigh length defined by $z_R=\frac{\pi w_0^2}{\lambda}$ where $w_0$ is the width of the beam and $\lambda$ is the centre wavelength \cite{tim}. The initial position of Bob corresponds to the position of the Rayleigh length and we assume that the detector has a width $\sqrt{2} w_0$ which captures all the intensity. Therefore, the transmission coefficient $t=1$ at this point. Away from the Rayleigh length, the transmission decreases with distance $L$ as:

\begin{equation}
t=t_0 \sqrt{\frac{2}{1+\left( \frac{L}{z_R} \right)^2}}
\end{equation}  

\begin{figure}
\centering
\includegraphics[width=1.12 \linewidth]{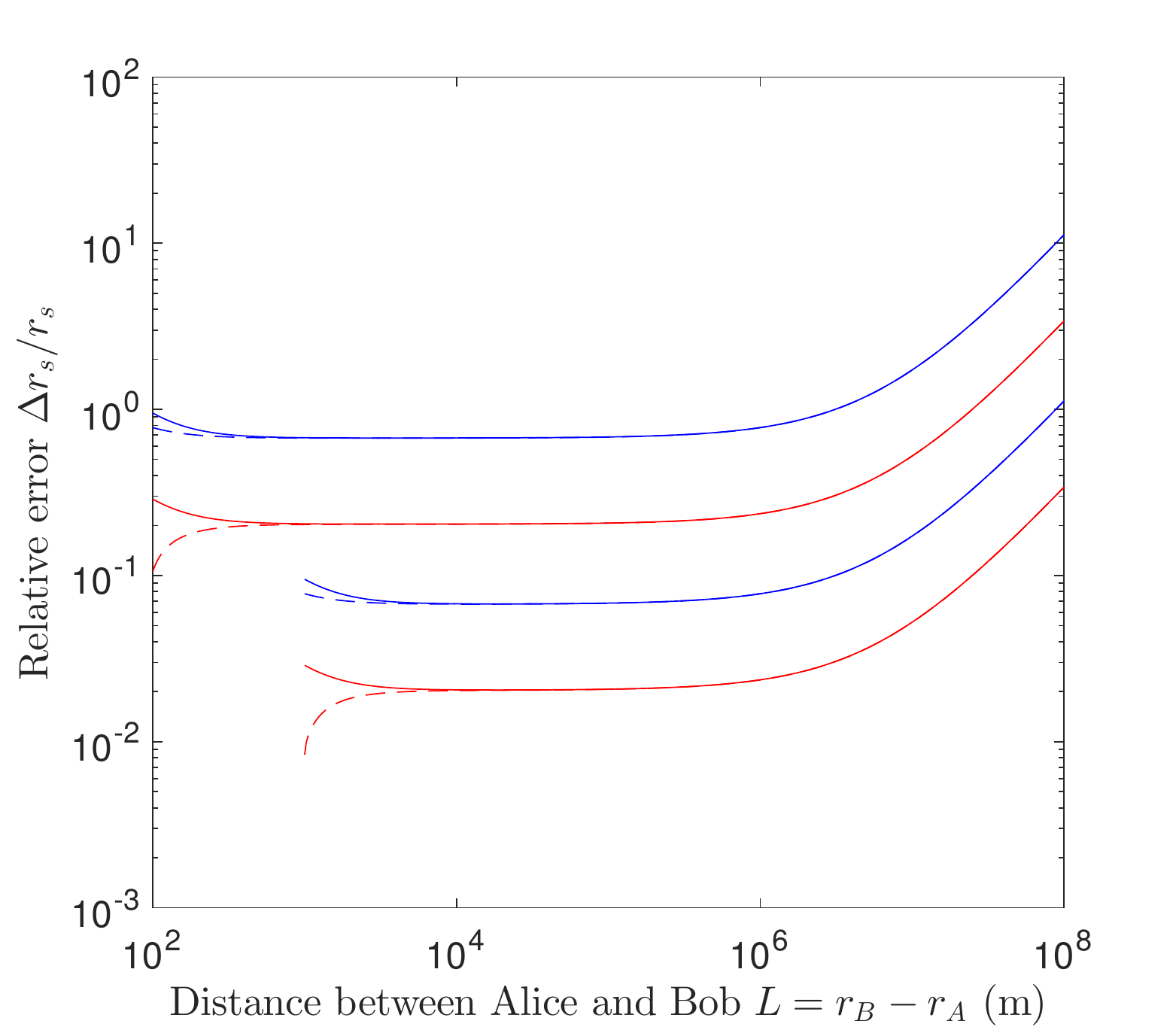}
\caption{Log- log plot of relative Schwarzschild error for Rayleigh lengths $z_R=100$m and $z_R=1000$m using Gaussian frequency profile (blue) at the optimal point $x=1/2$ and $\Delta=0.001$ rectangular frequency profile (red) also at its optimal point with $\tilde{n}=1000$. The solid lines are fully coherent photons with no squeezing, and the dashed lines have an added squeezing of 10 dB. (Other parameters: $N=200$ measurements, $r_A=6.37 \times 10^6$m, $r_B=42.0\times 10^6$ m, $\sigma=2000$, $\omega_0=700$ THz and hence $\delta=6.0 \times 10^{-10}$)}
\label{goodgauss}
\end{figure}


For the Gaussian frequency profile, the relative Schwarzschild error is plotted in Fig. \ref{goodgauss} for two Rayleigh lengths (blue curves). We compare between fully coherent photon probe states $\tilde{n}=1000$ (solid lines) and with additional squeezing of $10$ dB (dashed lines). With $10$ dB of squeezing, if Bob's distance is exactly at the Rayleigh length, the error is of the same order as the minima at $L\approx 10^5$ m. Nonetheless, the best situation is when $z_R=10^3$ m corresponding to a beam width of $w_0=1.4$ cm which has a minimum error of $10^{-1}$ between $L=10^4$ m and $L=10^6$ m. Squeezing has no effect on this minimum because the signal is heavily attenuated at Bob's location. For the Gaussian frequency profile, the best relative error is rather poor. We now present the $\tanh$ rectangular frequency profile as an alternative to this resource intensive scheme.

Now consider the red curves in Fig. \ref{goodgauss}, which report the relative error for the same parameters using a rectangular frequency profile with $\Delta=0.001$. We immediately note for the squeezed coherent probe state that there are no minima at any location. The minimum occurs at the Rayleigh lengths (where $t=1$) with an impressive 1 order of magnitude improvement in precision over the Gaussian frequency profile. Thus, it is now possible to measure the Schwarzschild radius with excellent precision with Bob at the Rayleigh lengths $L=z_R=100$ m and $L=z_R=1000$ m to achieve $10^{-1}$ and $10^{-2}$ relative error respectively. We note that squeezing has a significant effect at the Rayleigh length because $\Theta\approx 1$ since $L$ is small and the transmission coefficient is $t=1$. In this regime, squeezing becomes important. We observe that the precision increases up to a factor of $5 \times 10^{-1}$ for both Rayleigh lengths with the added $10$ dB of squeezing. However, for fully coherent photons, the minima are recovered and occur at approximately $L=10^5$ m. Therefore, a $\Delta=0.001$ rectangular frequency profile works well for short distances and are always up to 10 times more precise than Gaussian profiles of the same frequency spread.   

\subsection{Measurement basis of $r_s$}
In our derivation, the Bures distance definition of the QFI assumes a Gaussian measurement basis. To determine the measurement basis, we refer back to the definition of the QFI used by Ref \cite{MON07}. The eigenspace of $\hat \Lambda (\Theta)$ represents the optimal measurement for which the Fisher information is maximised. We note that the measurement strategy in Ref \cite{adaptive} takes this into account and a one-step adaptive strategy is proposed. In the first step, one makes a fraction of the total measurements $N^{\xi}$ where $1/2 < \xi < 1$ and provides an estimate for the parameter $\Theta_0$. Consequently, the eigenspace $\hat \Lambda (\Theta _0)$ can be built from the knowledge of $\Theta_0$ and used to make a better estimate on the remaining measurements. 

In Ref \cite{MON07}, it has been proven that the optimal measurement for the beamsplitter parameter using a pure Gaussian probe state is of the form $D(\alpha)S(r)S^{\dagger}(\eta)D(\beta)\ket{n}$. Where $D(\alpha)=\exp(\alpha a^{\dagger}-\alpha^* a)$ is the displacement operator and $S(r)=\exp(\frac{1}{2} r^2 a^{\dagger 2}-\frac{1}{2}r^{*2}a^2)$ is the squeezing operator. The coherent amplitude $\beta$ and squeezing parameter $\eta$ depends on the final evolved state. The measurement bases for the beamsplitter parameter using a lossy probe state are Gaussian operations and photon counting \cite{MON07}.

\section{Conclusion}

We have studied the optimal estimation of the beamsplitter parameter using a Gaussian quantum probe that is mixed due to loss. We determined the relevant Quantum Fisher informations using the definition of the Bures distance and an expression for the fidelity in terms of the quadrature variances. Using this convenient expression, we've considered a squeezed coherent probe state prepared by Alice. We have optimized the QFI with respect to the squeezing fraction and found that for very lossy probe states or low beamsplitter transmission, coherent states are more favourable. However, for low loss and high beamsplitter transmission, squeezing the photons is more advantageous. We've applied these results to a situation where loss is inevitable. In particular we've studied the use of a lossy probe state to estimate the Schwarzschild radius $r_s$ of Earth. We've shown that the frequency profile of the modes sent by Alice is important. By identifying the optimal point that Bob can choose to achieve the best precision, we determined that an approximate rectangular frequency profile achieves error bounds an order of magnitude better than a Gaussian. We also considered more realistic scenarios in which the transmission coefficient depends on the distance from the source and found that approximate rectangular mode shapes are better overall. 

The current analysis is restricted to Gaussian states. Other non-classical probe states such as NOON and entangled coherent states (ECS) may enhance the precision even further. Also, the current analysis assumes a specific protocol for estimating $r_s$. A comparison with other strategies would be interesting. 


\subsection*{Acknowledgements}
We would like to thank Daiqin Su and Antony Lee for useful discussions. This research was supported in part by the Australian Research Council Centre of Excellence for Quantum Computation and Communication Technology.

\appendix
\section{Mode Splitter}

Consider that we have an input mode that can be written:
\begin{equation}
a' = \sqrt{\epsilon} \; a + \sqrt{1-\epsilon} \; a''
\end{equation}
For our purposes the mode $a'$ can be considered the mode that Alice sent, as it appears when it reaches Bob, whilst $a$ is the mode that Bob expects. The mode $a''$ is the unmatched part which has the property $[a, a''^{\dagger}] = 0$. Now Bob applies a mode sensitive beamsplitter described by the unitary:
\begin{equation}
U = \exp{i \theta (a b^{\dagger} + b a^{\dagger})}
\end{equation}
This looks like a normal beamsplitter but it isn't because it only acts specifically on the modes $a$ and $b$ (where the vacuum mode entering the other beamsplitter port can be assumed to be in the mode $b$ without loss of generality). The transfer functions for the modes through the beamsplitter can be written, for the transmitted beam:
\begin{equation}
a'_T = \sqrt{\epsilon} (\sqrt{\eta} \; a + \sqrt{1-\eta} \; b) + \sqrt{1-\epsilon} \; a''
\end{equation}
where $\sqrt{\eta} = \cos{\theta}$. For the reflected beam:
\begin{eqnarray}
a'_R &=& \sqrt{1-\eta} \; a - \sqrt{\eta} \; b \nonumber \\
&=& \sqrt{1-\eta} (\sqrt{\epsilon} \; a' + \sqrt{1-\epsilon} \; v') - \sqrt{\eta} \; b
\end{eqnarray}
where in the second line we have introduced a vacuum mode defined as $v'= \sqrt{1-\epsilon} \; a - \sqrt{\epsilon} \; a''$. Notice if we set $\theta = \pi/2$ and hence $\eta =0$ we get the transformation we desire, i.e. $a'_R = \sqrt{\epsilon} \; a' + \sqrt{1-\epsilon} \; v'$ where $\sqrt{\epsilon} = [a, a'^{\dagger}]$.

One way to implement the mode-sensitive beamsplitter of Eq. 2 is described in Ref \cite{ECK11}.

\clearpage{}

\end{document}